\documentclass[fleqn,usenatbib]{mnras}

\usepackage{newtxtext,newtxmath}

\usepackage[T1]{fontenc}
\usepackage{ae,aecompl}

\usepackage{graphicx}	
\usepackage{amsmath}	
\usepackage{amssymb}	
\usepackage{natbib}
\usepackage[caption=false]{subfig}
\usepackage{ragged2e}
\usepackage{placeins}
\usepackage{hyperref}
\usepackage{bigints} 
\usepackage{physics}
\usepackage{xcolor}

\def\mean#1{\left< #1 \right>}

\title[Turbulence in the ICM]{Turbulence in the intracluster medium: simulations, observables \& thermodynamics}

\author[R. Mohapatra  \& Prateek Sharma]{
Rajsekhar Mohapatra,$^{1,2}$\thanks{E-mail: rajsekhar.mohapatra@anu.edu.au}
Prateek Sharma,$^{1,3}$\thanks{E-mail: prateek@iisc.ac.in}
\\
$^{1}$Department of Physics, Indian Institute of Science, Bangalore, 560012, India\\
$^{2}$ Research School of Astronomy and Astrophysics, Australian National University, Canberra, ACT 2611, Australia \\
$^{3}$ MPI f\"ur Astrophysik, Karl-Schwarzschild str. 1, Garching 85741, Germany \\
}

\date{Accepted XXX. Received YYY; in original form ZZZ}

\pubyear{2018}

\begin{document}
\label{firstpage}
\pagerange{\pageref{firstpage}--\pageref{lastpage}}
\maketitle

\begin{abstract}
We conduct two kinds of homogeneous isotropic turbulence simulations relevant for the intracluster medium (ICM): 
(i) pure turbulence runs without radiative cooling; (ii) turbulent heating$+$radiative cooling runs with global thermal balance.
For pure turbulence runs in the subsonic regime, the rms density and surface brightness (SB) fluctuations  vary as 
the square of the rms Mach number ($\mathcal{M}_{\text{rms}}$). However, with thermal balance, the density 
and SB fluctuations $(\delta SB/SB)$ are much larger.  These scalings  have
implications for translating SB fluctuations into a turbulent velocity, particularly for cool cores. 
For thermal balance runs with large (cluster core) scale driving, both the hot and cold phases of the gas are supersonic. 
For small scale (one order of magnitude smaller than the cluster core) driving, multiphase gas forms on a much longer timescale but $\mathcal{M}_{\text{rms}}$
is smaller. Both small and large scale driving runs have velocities larger than the Hitomi results from the Perseus cluster. Thus turbulent heating as the 
dominant heating source in cool cluster cores is ruled out if multiphase gas is assumed to condense out from the ICM. 
Next we perform thermal balance runs in which we partition the input energy into thermal and turbulent parts and tune their relative magnitudes. 
The contribution of turbulent heating has to be $\lesssim 10\%$ in order for turbulence velocities to match Hitomi observations.
If the dominant source of multiphase gas is not cooling from the ICM (but say uplift from the central galaxy), the importance of turbulent heating cannot be excluded.
\end{abstract}



\section{Introduction}\label{sec:Introduction} 
The intracluster medium (ICM) refers to the hot ($\sim 10^7-10^8$ K) X-ray emitting plasma that pervades clusters of galaxies. It contains majority of the baryons within the cluster. 
It is mainly composed of ionised hydrogen and helium, but also contains other elements such as iron. It loses energy via bremsstrahlung and metal line emission. 

The radiative cooling time is shorter for a higher density. Since the gas density is higher toward the cluster center, inner regions are expected to cool much faster than the outskirts. 
In relaxed cool core clusters the core is expected to cool, lose pressure support, and flow toward the center of the cluster (see \citealt{fabian1994cooling} for a review). The cooling gas is expected to 
cool all the way to form molecules and hence lead to active star formation. The cooling-only model predicts a star formation rate (SFR) $\sim$100-1000M$_\odot/$year in cool core clusters. 
However, observations show a much reduced SFR (by orders of magnitude; e.g., \citealt{odea2008}). This is known as the cooling flow problem.

It is now accepted that the cool cores lose thermal energy due to radiative cooling, but most of the losses are compensated by heating due to other sources such as thermal conduction, 
cosmic rays, and turbulence. Heating due to active galactic nuclei (AGN) jets powered by accretion on to the central supermassive 
black hole (SMBH) is particularly attractive because of sufficient energy and negative feedback (see \citealt{mcnamara2007} for a review). 
A cool, dense core is prone to condensation of cold gas in the core that enhances accretion on to the SMBH
and the jet power. A much larger jet power driven by multiphase condensation is able to stop catastrophic cooling in the core and the cycle continues (e.g., see \citealt{prasad2015,li2015}). 

Cool cluster cores show multiphase gas (at $\sim 10$ K traced by CO [e.g., \citealt{edge2001}], at $\sim 10^4$ K traced by nebular lines [e.g., \citealt{hu1992,mcdonald2012}], 
and of course the diffuse ICM at $10^7-10^8$ K). The multiphase
gas can be interpreted in terms of local thermal instability in an ICM with global thermal balance (\citealt{sharma2010thermal,sharma2012thermal}).
The feedback model proposes that heating by AGN jets  
acts like a time-delayed feedback loop, which injects on-average the same amount of energy lost via cooling back into the ICM through energetic outbursts 
\citep{rafferty2006feedback,prasad2015}.

Energy injection through AGN feedback and sloshing of the ICM during mergers (mostly with small subhalos) are expected to drive motion in the ICM. 
Turbulent structures, density and pressure fluctuations, have been observed in the ICM (\citealt{schuecker2004,zhuravleva2014turbulent,khatri2016}). Turbulence has 
been proposed as a mechanism through which AGN jets and mergers can heat the ICM via direct turbulent heating (\citealt{zhuravleva2014turbulent}; but see 
\citealt{falceta-goncalves2010,bambic2018}) or via mixing of the much hotter outskirt/bubble gas 
with the ICM (e.g., \citealt{banerjee2014turbulence,hillel2017}). From the Kolmogorov (hereafter, K41) picture 
of homogeneous, isotropic turbulence 
\citep{kolmogorov1941dissipation}, turbulent energy from the driving scale cascades down the length scales before being dissipated at the viscous scale, thus heating 
the ICM. Other than heating, turbulence also plays two opposing roles in multiphase condensation: it can generate large density fluctuations, thus aiding condensation of cold filaments; 
it can mix up the cooling gas with the hot phase, thereby inhibiting multiphase condensation.

Two recent observational studies -- \citet{aharonian2016quiescent} (the Hitomi collaboration) and \citet{zhuravleva2014turbulent} -- obtain a similar estimate for the turbulent velocities
in the  core of Perseus cluster.
While \citet{zhuravleva2014turbulent} reconstruct the velocity amplitudes by analysing the power spectrum of X-ray surface brightness fluctuations, Hitomi directly 
measured the line of sight velocity dispersion ($\sigma_{\text{LOS}}$) by analysing the broadening of Fe {\tiny XXV} and Fe {\tiny XXVI} lines. \citet{zhuravleva2014turbulent} 
find the turbulent energy injection to be large enough to completely compensate  radiative cooling losses. On the other hand, the Hitomi paper emphasizes that the ICM is quiescent, 
and the turbulent pressure is only $4\%$ of the thermal pressure. Of course, even such a small turbulent velocity can be sufficient to check radiative cooling in the core,
provided that the driving scale of the turbulence is sufficiently small (but see \citealt{bambic2018};\footnote{\citet{bambic2018} argue that the time for turbulence to travel to the entire
cool core is longer than the cooling time.
Another interpretation of this argument is that the turbulent
heating rate $\rho v_L^3/L$ can be large if the driving scale $L$ is sufficiently small. But if $L$ is too small compared to the core size, turbulence needs to be driven independently throughout
the core because energy primary cascades to small scales in Kolmogorov turbulence.} 
see also the first bullet-point in section \ref{sec:conclusion}).

Thus, some of the unanswered questions are: what fraction of ICM feedback heating can be due to turbulent dissipation; the source of cold gas -- whether most of 
it is uplifted or cooling down from the hot ICM; and whether the observed density perturbations are indeed 
generated by stratified turbulence (an assumption underlying the treatment of \citealt{zhuravleva2014turbulent}; density perturbations can also arise from the local thermal 
instability, leading to the separation of hot and cold phases of gas without generating much turbulence). Although we focus on turbulence-driven density 
perturbations in cool-core clusters, we also briefly discuss pressure fluctuations that can be probed 
by the fluctuating Sunyaev-Zeldovich signal out to the virial radius (\citealt{khatri2016}). Thus, our results on isotropic/homogeneous turbulence are also applicable to non-cool-core 
clusters and the circumgalactic medium, particularly at small scales.

In subsonic K41 turbulence, the velocity and density fluctuations at a particular length scale $l$ scale as $l^{1/3}$ $(v_l$, $\delta\rho_l \propto l^{1/3}$; 
\citealt{kolmogorov1941dissipation,corrsin1951spectrum}). This is because the turbulent energy cascade rate $\epsilon$ is a constant in the inertial regime, given by 
 $\epsilon = \rho v_l^3/l$, and the density behaves like a passive scalar mixed by turbulent eddies. For subsonic turbulence, density variations are small. From these simple scaling relations, we find $v_l\propto l^{1/3}$. Density fluctuations 
 follow the same scaling as $v_l$; therefore, $\delta \rho_l \propto l^{1/3}$. In steady state, on average, this cascading rate 
 $\epsilon$ is  the rate at which turbulent energy is injected into the system at the driving scale and the rate at which it is dissipated at the viscous scale.

In an earlier numerical study \citet{banerjee2014turbulence},which assumed that the majority of cold gas in cluster cores is due to condensation from the ICM, 
showed that when turbulent heating rate ($\rho v_l^3/l$) balances radiative cooling rate, the required turbulent velocities 
are sonic (Mach number close to unity). But
cool cores are known to be subsonic. This study assumed the turbulence driving scale to be $\sim 10$ kpc, comparable to the size of the cool core and AGN bubbles/X-ray cavities. However, 
if we decrease the driving scale while still maintaining the same energy injection rate $\epsilon$, we can decrease $v_l$ since $v_l \propto l^{1/3}$ for a constant $\epsilon$. This way
we can still achieve subsonic velocities driven by turbulent forcing, while still maintaining the global thermal balance between radiative cooling and turbulent heating. But driving turbulence 
at smaller length scales would also lead to smaller density fluctuations, since $\delta \rho_l \propto v_l \propto l^{1/3}$. 
Turbulence driven at smaller scales not only drives weaker turbulence-driven density perturbations, but also suppresses the mixing of hotter and cooler phases at large scales. 
In this paper we study the impact of the driving scale on the turbulence in cool cluster cores.

Imposing thermal balance between turbulent heating and radiative cooling, $\rho v_L^3/L \sim n^2 \Lambda$ ($L$ is the driving scale, $n$ is electron/ion number density and $\Lambda[T]$
is the cooling function) and assuming the expected scalings with the halo mass, implies that the Mach number of the largest eddies ${\cal M} \propto (nL \Lambda)^{1/3}/c_s$ 
($c_s$ is the sound speed of the ICM) is rather insensitive to the halo mass. Additionally, if we assume 
that majority of the observed cold gas in cool cluster cores is produced as a result of cooling from the hot phase (this is plausible but not at all an established fact), then the cooling 
time of the cooling blob must be shorter than the turbulent 
mixing time. This condition constraints the Mach number in the hot phase to be larger than a threshold value, which is larger than unity if turbulent heating is the dominant heating source and 
driving occurs at the core scale (c.f. Eq. \ref{eq:cool_mix}). Another possibility is that the cold gas in cluster cores is not due to condensation from the hot phase, but say because of uplift by 
AGN jets and buoyant bubbles (e.g., \citealt{revaz2008}). In this case the cooling time of the hot phase can be much longer than the turbulent mixing time and the Mach number in the hot phase
can be smaller than unity (at least for small [large] enough $L~[c_s]$). However, this possibility does not naturally explain the occurrence of multiphase gas only in clusters with the ICM density 
larger than a certain threshold (e.g., \citealt{cavagnolo2008}).

In the first part of our study,  we simulate homogeneous isotropic turbulence to derive the relations between gas density (pressure), surface brightness (projected pressure) fluctuations, and 
the turbulent Mach number of the flow (see section \ref{sec:ResultsT}).  
In the second set of runs, more applicable to cool cores, we impose global thermal balance over the entire simulation domain. In these runs, we analyse the thermodynamics of the flow, through 
Mach number and temperature distribution of the gas. We check the dependence of these thermodynamic aspects on the driving scale, fraction of turbulent heating relative to cooling, and initial 
density perturbations. Results from these simulations are presented in \autoref{sec:ResultsHC}.

In \autoref{sec:discussion} we present the caveats of our setup, and discuss our thermal balance results in the context of X-ray surface brightness fluctuations in cool cores and 
the 1-D line of sight velocity dispersion as measured by Hitomi in the core of Perseus cluster. We conclude in section \autoref{sec:conclusion}.

\section{Methods}\label{sec:methods} 
\subsection{Model equations}
We model the ICM using the hydrodynamic equations. As the ICM plasma is hot and fully ionised, the magnetic fields can have significant effects. 
From Alfv\'en's flux freezing theorem, field lines are frozen 
into the plasma and have to move along with it. In addition, the microscopic transport of heat along magnetic fields can lead to new buoyancy
instabilities (\citealt{balbus2000,quataert2008}) and enhanced mixing in galaxy clusters (\citealt{sharma2009,kannan2017}). 
Note that the kinetic whistler instability may significantly 
suppress thermal conduction (\citealt{levinson1992,roberg-clark2016}).

The aim of the present paper is to study the interplay of turbulence, cooling and density perturbations. \citet{banerjee2014turbulence} show that for our setup the evolution of 
magnetohydrodynamic (MHD) equations with anisotropic thermal conduction gives results qualitatively similar to the hydro simulations. The kinetic energy density in MHD is roughly half that in hydro 
(see their Fig. 1) and the density fluctuations are larger by a similar factor (see their Fig. 4). Because of smaller turbulent velocities the temperature and Mach number distributions in MHD 
are more bimodal (see their Figs. 6 \& 7). As expected, thermal conduction tends to wipe out small scale structure (\citealt{gaspari2013constraining}).
Since the overall impact of magnetic fields on thermodynamics and dynamics of the high beta ICM is easy to understand qualitatively, evolving HD equations is reasonable for our purposes. 

We model the core of the ICM using periodic boundary conditions, ignoring the shallow gradients in density and temperature. 
Since we model gas at high temperatures 
($T\gtrsim 10^4K$), we ignore self-gravity in our simulations.  
We solve the following equations:
\begin{subequations}
\begin{align}
\label{eq:Euler1}
&\frac{\partial\rho}{\partial t}+\nabla\cdot (\rho \mathbf{v})=0,\\
\label{eq:Euler2}
&\frac{\partial(\rho\mathbf{v})}{\partial t}+\nabla\cdot (\rho \mathbf{v}\otimes \mathbf{v})+\nabla P=\mathbf{F},\\
\label{eq:Euler3}
&\frac{\partial E}{\partial t}+\nabla\cdot ((E+P)\mathbf{v})=\mathbf{F}\cdot\mathbf{v}+Q-\mathcal{L},
\end{align}
\end{subequations}
where $\rho$ is the gas mass density, $\mathbf{v}$ is the velocity, $P=\rho k_B T/(\mu m_p)$ is the pressure, $\mathbf{F}$ is the turbulent force per unit volume that we apply, $E=\rho v^2/2 + P/(\gamma-1)$ 
is the total energy density, $\mu$ is the mean molecular mass, $m_p$ is the proton mass, $k_B$ is the Boltzmann constant, $T$ is the temperature, $Q(t)$ and $\mathcal{L}(\rho,T)$ are the thermal heating and cooling 
rate densities respectively, and $\gamma=5/3$ is the adiabatic index. The cooling rate density $\mathcal{L}$ is given by
\begin{equation}
\mathcal{L} = n_en_i\Lambda(T),
\end{equation}
where $\Lambda(T)$ is the temperature-dependent cooling function of \citet{Sutherland1993} corresponding to $1/3^{\text{rd}}$ solar metallicity, and $n_e$ and $n_i$ 
are electron and ion number densities respectively. 
The turbulent forcing $\mathbf{F}$ is applied using a spectral forcing method, as described in \autoref{subsec:Turb_forcing}.  In some runs, we include uniform thermal heating ($Q$)
throughout the domain, such that cooling is balanced by the sum total of turbulent and thermal heating. Viscosity and thermal conduction are not included explicitly. 

We carry out two sets of simulations (see Tables \ref{tab:turb_runs} and \ref{tab:cool_runs}). 
In turbulence-only simulations we vary the forcing amplitude to drive turbulence at different Mach numbers, but do not include cooling. The aim of these 
runs is to relate density (pressure) and surface brightness (projected pressure) fluctuations to the turbulent velocity for isotropic/homogeneous turbulence relevant 
below the Ozmidov scale (the scale at which the internal gravity wave oscillation timescale equals the turbulent eddy timescale; \citealt{ozmidov1965}). 
In the second set of simulations 
we impose thermal balance averaged over the whole computational domain to mimic the observed global thermal balance in cool cluster cores; i.e., the sum of the 
work done per unit time by turbulent forcing and the thermal power input equals the volume integrated cooling rate. In the thermal balance runs the denser/cooler regions cool and the 
hotter regions are heated (slowly) by design. Thus the temperature of the hot phase increases steadily and the CFL time step becomes shorter, making the second set of runs 
more time consuming. With gravity the hot regions will rise and cooler blobs will sink, but this physics is not included for the simulations in this paper. Our simulations are thus more relevant for scales below the Ozmidov scale, below which the Richardson number $\rm Ri\lesssim 1$ and turbulence dominates over buoyancy effects (see \autoref{eq:Richardson} for the definition of $\rm Ri$.)

\subsection{The cooling cutoff}
\label{subsec:coolcutoff}

In the absence of a gravitational field (and consequent stratification), cold gas can separate out from the hot phase due to local thermal instability. This cold gas collapses to an extremely small scale (\citealt{field1965thermal,koyama2004field,sharma2010thermal}). In order to prevent the cold gas from collapsing to an extremely small scale we cut off the cooling function at a 
temperature $T_{\rm cutoff}$. The scale of collapsing clouds, assuming isobaric
conditions, goes as $T_{\rm cutoff}^{1/3}$. For a very short cooling time, the isobaric condition is not valid during collapse
and the gas can fragment on the scales of $c_s t_{\rm cool}$  (\citealt{mccourt2018})
where 
\begin{equation}
\label{eq: cs}
c_s \equiv  \left ( \frac{\gamma P}{\rho} \right)^{1/2} = \left ( \frac{\gamma k_B T}{\mu m_p} \right)^{1/2} 
\end{equation}
is the sound speed and 
\begin{equation}
t_{\rm cool} \equiv \frac{3}{2} \frac{n k_BT}{n_e n_i \Lambda}
\end{equation}
is the cooling time evaluated at the temperature of the cold stable phase.

To prevent cold gas from collapsing to unresolvable small scales, we truncate  the cooling function at a small temperature floor $T_{\text{cutoff}}$. Thus, the cooling rate now has a form
\begin{equation}
\mathcal{L}=n_en_i\Lambda(T)\mathcal{H}(T-T_{\text{cutoff}}),
\end{equation}
where $\mathcal{H}$ is the Heaviside function. We choose $T_{\text{cutoff}} = 10^6\ K$ for most runs (we also tried a few runs with $T_{\text{cutoff}}=10^4\ K$ to check the sensitivity to this parameter). This choice is reasonable, since most of the gas that cools to $10^6$ K will cool to $10^4$ K because of a short cooling time in this temperature range. A higher cutoff temperature
enables us to better resolve the cold phase.

\subsection{Turbulent forcing}
\label{subsec:Turb_forcing}

We follow a spectral forcing method using the stochastic Ornstein-Uhlenbeck (OU) process to model the turbulent force $\mathbf{F}$ with a finite autocorrelation time scale $\tau$ (\citealt{eswaran1988examination,schmidt2006numerical}). 
The acceleration in the Fourier space is given by
\begin{equation}
\mathbf{a}_{\mathbf{k}}^{n}=f\ \mathbf{a}_{\mathbf{k}}^{n-1}+\sqrt{1-f^2}\ \mathbf{a^\prime}_{\mathbf{k}}^{n},
\end{equation}
where the exponential damping factor $f=\exp(-\delta t_n/\tau)$ ($\delta t_n$ is the $n^{\rm th}$ time step size), 
$\mathbf{a^\prime}_{\mathbf{k}}^{n}$ is the $n^{\text{th}}$ acceleration amplitude generated by our random number generator, 
$n$ being a time-step label. It is generated by a Gaussian random number generator with amplitude $A_{\text{turb}}$. 
We make sure that the driving acceleration is solenoidal, by subtracting its component along $\mathbf{k}$ in Fourier space, and taking only the solenoidal component,
\begin{equation}
\mathbf{a}_{\mathbf{k}}^n=\mathbf{a}_{\mathbf{k}}^n-\frac{\mathbf{a}^n_{\mathbf{k}}\cdot \mathbf{k}}{|\mathbf{k}|^2}\mathbf{k}.
\end{equation}
We limit the modes to which forcing is applied in the Fourier space by setting two limits $k_{\text{min}}$ and $k_{\text{max}}$, which control the distribution of $\mathbf{F}_k$ in the Fourier space. The force $\mathbf{F}^n(\mathbf{x})$ in the real space is given by:
\begin{equation}
\mathbf{F}^n(\mathbf{x})=\rho(\mathbf{x})\Re{\bigintss_{-\infty}^{\infty}\left(\sum_{|k|=k_{\text{min}}}^{k_{\text{max}}}\mathbf{a}_{\mathbf{k}}^n\right)e^{-\iota \mathbf{k}\cdot\mathbf{x}} \text{d}\mathbf{x}}.
\end{equation}
The typical values of $k_{\text{min}}$ and $k_{\text{max}}$ are of the order of $2\pi/(10\ \text{kpc})$. We label the wavenumbers $\mathbf{k} = 2 \pi \mathbf{K}/L$ by $\mathbf{K}~(L$ is the box-size) that are 
indicated in \autoref{tab:turb_runs} and \autoref{tab:cool_runs} for each run. 

We make sure that turbulent forcing does not add any net momentum to the computational box. We subtract a constant from all three components of momentum at all grid points, such that $\mean{\rho \delta\mathbf{v}(\mathbf{x})}=0$ at each time step ($\langle \rangle$ denotes volume average and $\delta\mathbf{v}$ is the change in velocity at a grid point due to turbulent forcing).

In heating balancing cooling runs, we scale turbulent forcing $\mathbf{F}^n$ so as to maintain global thermal equilibrium, i.e., we explicitly enforce the following condition:
\begin{equation}
\mean{\mathbf{F}\cdot (\mathbf{v}+\delta\mathbf{v})}+Q=\mean{\mathcal{L}}.
\end{equation}
We introduce a parameter $f_{\text{turb}}$ which denotes the fraction of cooling that is compensated by turbulent heating. To maintain thermal balance, the gas is thermally heated uniformly at a rate $Q=(1-f_{\text{turb}})\mean{\mathcal{L}}$ at each grid point.

\subsection{Initial density perturbations}\label{subsec:Density-perturbations}
In some of our runs, we initialize isobaric density fluctuations on top of the uniform density, which are generated such that $\rho_k$, the Fourier transform of $\rho$, has a scaling similar to that
expected in a steady turbulent flow, i.e., $\rho_k \propto k^{-1/3}$. Our initial density fluctuations follow this scaling for $\sqrt{2} \leq K \leq 12$.
 
\subsection{Numerical Methods}\label{subsec:numericalmethods}
We use a modified version of the grid based PLUTO code (version 4.1; \citealt{mignone2007pluto}) for our simulations. 
We evolve the Euler equations in PLUTO, with additional forcing, cooling and heating terms added as source terms (Eqs. \ref{eq:Euler1}-\ref{eq:Euler3}). The relation between pressure, density and temperature is set by the ideal gas equation of state. We use the tvdlf (Total Variation Diminishing Lax-Friedrich) solver, with periodic boundary conditions, RK-3 time stepping and parabolic reconstruction. All our runs use a box size of 40 kpc in each direction, with 3D Cartesian grids having a resolution of $256^3$. We have tested the code for 
numerical convergence by doubling the resolution to $512^3$ for some of the runs. Our box size of 40 kpc ensures that we are able to focus on the cool core 
and have a good resolution, upto 100 pc.

We initialize the gas with a temperature of $T_0=1.03$ keV, $n_e=0.1 \text{ cm}^{-3}$, which give a cooling time $\approx$ 60 Myr. We assume the gas composition to have $\mu=0.5$ and $\mu_e=1.0$ (although we use a cooling function corresponding to $Z_\odot/3$ metallicity). For our thermal balance runs, the local thermal instability leads to the separation of gas into hot and 
cold phases, and the rarer/hotter phase (which is hotter and rarer than the initial condition) represents the ICM.

\begin{table*}
	\centering
     \caption{Turbulence-only runs}
     \label{tab:turb_runs}
     \resizebox{\textwidth}{!}{
	\begin{tabular}{lcccc} 
		\hline
		Label & Resolution & Forcing Amplitude ($A_{\rm turb}$) & $K_{\text{driving}}$ & Remarks\\
		\hline
		Fl1 & $256^3$ & 0.005 & $0<K_{\text{driving}}\leq\sqrt{2}$ & subsonic\\
		Fl2 & $256^3$ & 0.02 & $0<K_{\text{driving}}\leq\sqrt{2}$ & subsonic\\
		Fl3 & $256^3$ & 0.1 & $0<K_{\text{driving}}\leq\sqrt{2}$ & transonic initially\\
		Fl4 & $256^3$ & 0.9 & $0<K_{\text{driving}}\leq\sqrt{2}$ & supersonic initially\\
		Fl5 & $256^3$ & 2.5 & $0<K_{\text{driving}}\leq\sqrt{2}$ & supersonic initially\\
		Flr & $512^3$ & 0.005 & $0<K_{\text{driving}}\leq\sqrt{2}$ & subsonic, results converge\\
		Fh & $256^3$ & 0.1 & $K_{\text{driving}}=12$ & subsonic \\
        \hline
        
	\end{tabular}}
    \justifying \\ \begin{footnotesize} In the labels, F stands for Fiducial (without explicit heating \& cooling), r denotes the high resolution run with $512^3$ grid points, l denotes driving at low 
    $k$s ($0< K_{\text{driving}}\leq\sqrt{2}$), h denotes driving at high $k$s ($K_{\text{driving}}= 12$). \end{footnotesize} 

\end{table*}
\section{Results -- Turbulence-only runs}\label{sec:ResultsT}
Here we describe the results of our fiducial (turbulence-only, no cooling or thermal heating) simulations. Table \ref{tab:turb_runs} lists the parameters of these runs. 
The energy equation (Eq. \ref{eq:Euler3}) is thus
\begin{equation}
\frac{\partial E}{\partial t}+\nabla\cdot ((E+P)\mathbf{v})=\mathbf{F}\cdot\mathbf{v}.
\end{equation}
The gas temperature increases with time due to work done by turbulent forcing, the strength of which we characterize by an amplitude $A_{\text{turb}}$.

The flow is subsonic for lower values of $A_{\text{turb}}$, and transonic/supersonic at early times for a large $A_{\text{turb}}$. Since we do not have radiative cooling in these runs, the gas eventually heats up, 
and the flow always becomes subsonic at later times as the sound speed $c_s$ increases. 

\subsection{Mach number and density/pressure perturbations}\label{subsec:rhomach}

\begin{figure}
	\includegraphics[width=\columnwidth]{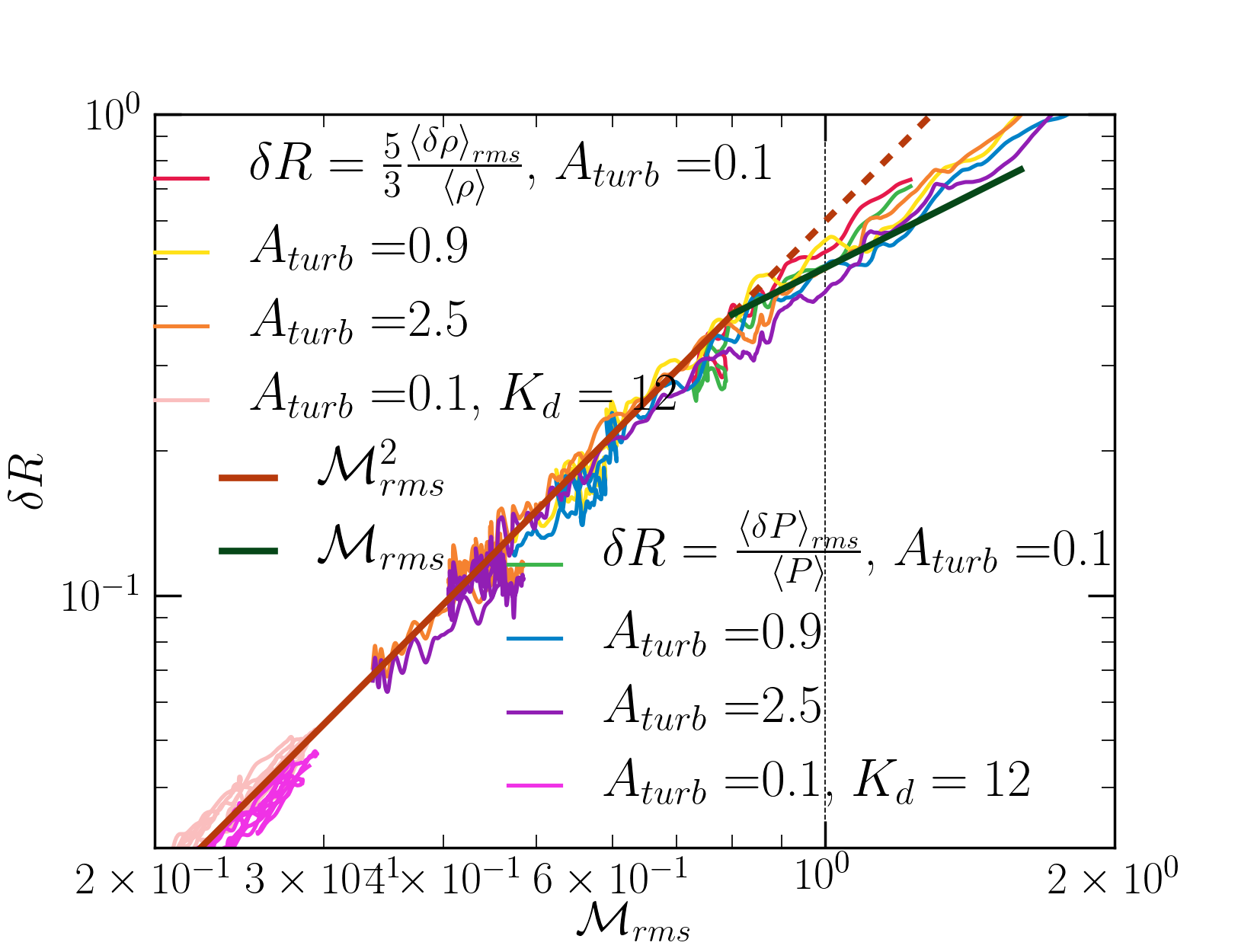}
    \caption[Density fluctuations vs Mach number]{The root mean square (rms) density and pressure fluctuations as a function of the rms Mach number $\mathcal{M}_{\text{rms}}$. 
    Both these fluctuations vary $\propto \mathcal{M}_{\text{rms}}^2$ in the subsonic regime and flatten in the supersonic regime. The 
    data are plotted after the first maximum in  $\mathcal{M}_{\text{rms}}$, roughly after the saturation of turbulence. The trends of pressure and density fluctuations are strikingly 
     similar. Also note that the strongest driven system achieves smaller rms Mach number and density/pressure fluctuations at a much faster rate. This is because the turbulent 
     heating time $3p/(2 \rho v_L^3/L) = 0.9 {\cal M}_{\rm rms}^{-2} (L/v_L)$ is much shorter than the eddy turnover time $(L/v_L)$ for a larger Mach number, where $L$ is the driving scale 
     and $v_L$ is the velocity at this scale.}
    \label{fig:rho-mach}
\end{figure}

\autoref{fig:rho-mach} shows the relative root mean square (rms) fluctuations in density $\mean{\delta\rho}_{\text{rms}}/\mean{\rho}$ and pressure 
$\mean{\delta P}_{\text{rms}}/\mean{P}$ as a function of the rms Mach number 
\begin{equation}
\label{eq:Mrms}
\mathcal{M}_{\text{rms}} \equiv \frac{ \mean{v_{\text{rms}}^2}^{1/2} }{ c_s},
\end{equation}
where $v_{\rm rms}$ is the rms velocity and $c_s$ is the volume-averaged sound speed. In the subsonic regime ($\mathcal{M}_{\text{rms}}<0.8$), 
the density and pressure fluctuations vary as $\mathcal{M}_{\text{rms}}^2$. In the transonic/supersonic regime, both density and pressure fluctuations flatten with 
$\mathcal{M}_{\text{rms}}$ (e.g., see Fig. 7 in \citealt{nolan2015}). Further, the density and pressure fluctuations 
scale linearly with each other in the subsonic regime but in the shock-dominated supersonic regime the density behind a shock can only be a factor of 4 higher than the ambient
value but the pressure can be much higher, indicating the breakdown of the linear scaling.
 
The scaling of density and pressure fluctuations with the Mach number can be motivated from the following arguments. In the subsonic regime, the flow is close to incompressible and the
pressure satisfies the Poisson equation $\nabla^2 P = \rho \nabla {\bf v} : \nabla {\bf v}$, which implies that $\delta P \sim \rho \delta v^2$, or 
$\delta P/P \sim \gamma \delta v^2/c_s^2 \sim \gamma \mathcal{M}_{\text{rms}}^2$. For transonic Mach numbers the disturbance are dominated more and more by sound-like 
perturbations with $\delta P \sim \rho c_s \delta v$ and $\delta P/P \sim \gamma \delta v/c_s \sim \gamma \mathcal{M}_{\text{rms}}$ (this is just the relation between 
fluctuations in a sound wave). In both subsonic and transonic regimes the 
pressure and density fluctuations are related as $\delta P/P \sim \gamma \delta \rho/\rho$. These scalings explain the observed relation in \autoref{fig:rho-mach}. The top-left panel in
Fig. 6 of \citet{konstandin2012} shows a similar scaling of density fluctuations\footnote{They measure $\sigma_\rho$ the width of the PDF of $\ln \rho$, which in the subsonic regime 
should roughly equal $\langle \delta \rho \rangle_{\rm rms}/\langle \rho \rangle$. We have 
done some low Mach number simulations with pure compressible driving, and find that $\langle \delta \rho \rangle_{\rm rms}/\langle \rho \rangle$ and $\langle \delta P \rangle_{\rm rms}/\langle P \rangle$ 
scalings are closer to $\propto {\cal M}_{\rm rms}^2$ than $\propto {\cal M}_{\rm rms}$ for ${\cal M}_{\rm rms} \gtrsim 0.2$. This does not agree with Fig. 6 of \citet{konstandin2012}, 
but it may be because they are using an isothermal
equation of state for which pressure is a constant times the density and our pressure fluctuations are governed by Eqs. \ref{eq:Euler1}-\ref{eq:Euler3}.} 
and Mach number as ours for their isothermal turbulence simulations with solenoidal driving (like us).

\begin{figure}
	\includegraphics[width=\columnwidth]{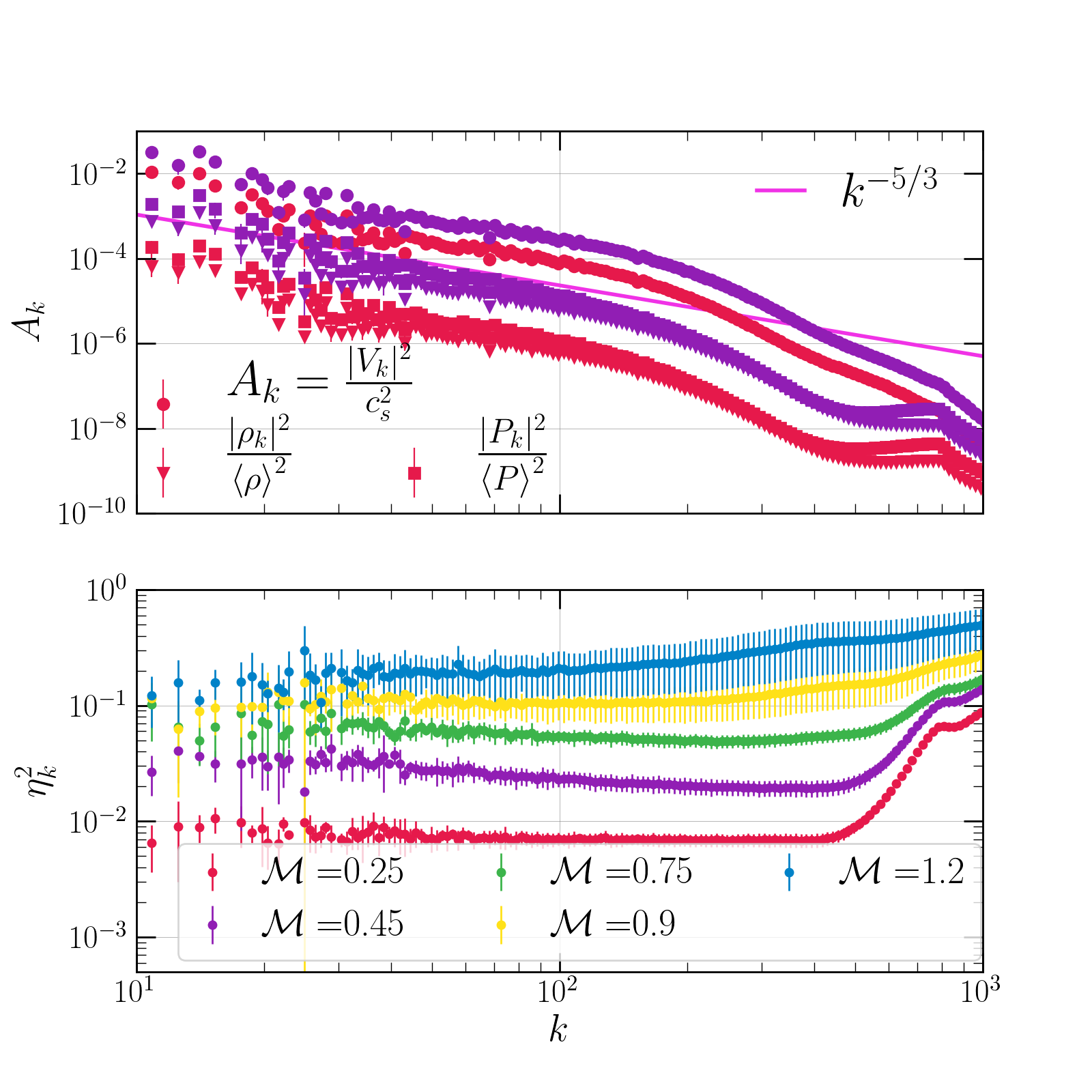}
    \caption[Velocity and density/pressure power spectra]{Upper panel: Power spectra of velocity, density and pressure for low Mach numbers relevant to the ICM ($\mathcal{M}_{\text{rms}} = 0.25,~0.45$).
    Density, pressure and velocity closely follow the K41 $k^{-5/3}$ scaling in the inertial range. 
    Lower panel: The ratio $\eta_k^2=\frac{\rho_k^2}{\left<\rho\right>^2}/\frac{V_k^2}{c_s^2}$ of density and velocity 
    power spectra has a significant flat region for subsonic flows. In the transonic regime ($\mathcal{M}_{\text{rms}} \gtrsim 1$) the density spectrum is steeper than the velocity spectrum.  
    Note that the ratio $\eta_k^2$ increases with $\mathcal{M}_{\text{rms}}$. The data for a given ${\cal M}_{\rm rms}$ are from different low$-k$ driving runs in \autoref{tab:turb_runs}.
    Error-bars in both panels (error-bars are not easily seen in the top panel) correspond to 1-$\sigma$ variation across the mean in different samples with the same ${\cal M}_{\rm rms}$. The variations 
    are larger for ${\cal M}_{\rm rms} \gtrsim 1$ in the bottom panel, reflecting higher variability with a larger Mach number.
    }
    \label{fig:rho-vel-spectra}
\end{figure}

\subsection{Power spectra}
\label{subsec:powspectra}

Now that we have established that the domain averaged rms density and pressure fluctuations vary as $\mathcal{M}_{\text{rms}}^2$ for subsonic turbulence relevant 
for the ICM, we move on to power spectra. We find that the spectral amplitudes of both velocity and density perturbations $\rho_k$ and $v_k$ vary as $k^{-1/3}$ (i.e.,
density follows the Obukhov-Corrsin spectrum for passive scalars; \citealt{corrsin1951spectrum}). Although the density and velocity power spectra have the same slope, from \autoref{fig:rho-mach} and
Parseval's theorem (equal power in real and Fourier space) we expect $\delta \rho_k/v_k$ to increase with an increasing 
$\mathcal{M}_{\text{rms}}~(\delta \rho_k/\mean{\rho}\propto \mathcal{M}_{\text{rms}}\delta v_k/c_s)$.  
\autoref{fig:rho-vel-spectra} shows the density/pressure and velocity power spectra (top panel) and their ratio (bottom panel) for some of our fiducial runs. Notice 
 a large flat portion in the ratio between power spectra of density and velocity, but an increasing value of the ratio with an increasing Mach number. 

The top panel of  \autoref{fig:rho-vel-spectra} shows that the density and pressure power spectra are very similar for ${\cal M}_{\rm rms} \gtrsim 0.25$.  We expect the pressure spectrum to be steeper by unity
than the density spectrum (which follows the passive-scalar/velocity spectrum) in the very subsonic regime (e.g., see Eq. 6.94 in \citealt{lesieur1997turbulence}). However, for the Mach numbers relevant
for galaxy clusters we find an almost the same spectral slope for the pressure and density power spectra, with only a slight hint of steepening of the former at the smallest Mach numbers.

The ratio of the density and velocity power spectra 
is proportional to $\epsilon_{\rho}/\epsilon_{v}$, where $\epsilon_\rho$ is the density fluctuation flux and $\epsilon_v$ is the kinetic energy flux (both are constant in the inertial range), 
which can be defined as
\begin{align}
&\epsilon_{\rho}(l)=\frac{\delta\rho(l)^2v(l)}{l},\\
&\epsilon_{v}(l)=\frac{v(l)^3}{l},\\
& \eta(l)^2 \equiv \frac{c_s^2}{v(l)^2} \frac{(\delta \rho(l))^2}{\mean{\rho}^2} =\frac{\epsilon_{\rho}(l)}{\epsilon_{v}(l)} \frac{c_s^2}{\mean{\rho}^2},
\end{align}
where $v(l)$ is the characteristic velocity at length scale $l$ (note that the labels $l$ and $k$ are interchangeable). 
The ratio of the power spectra is constant in the inertial range as seen in \autoref{fig:rho-vel-spectra}, and $\epsilon_v$ , $\epsilon_{\rho}$ are constants independent of $l$. 
Note that these arguments need to be modified for transonic/supersonic turbulence. In the transonic runs, the inertial range is not flat, and \autoref{fig:rho-vel-spectra} shows that there 
is a slight increase in the ratio $\eta_k^2 \equiv \frac{\rho_k^2}{\left<\rho\right>^2}/\frac{V_k^2}{c_s^2}$ with an increase in $k$. Perhaps most importantly for the ability to convert density 
fluctuations to turbulent velocities, the ratio of powers in density and velocity perturbations at a given scale is proportional to the Mach number. This is found to be a constant
in previous works that include stratification (\citealt{zhuravleva2014relation,gaspari2014}).

\subsection{Surface brightness fluctuations}
\label{subsec:sb}

In X-ray observations we directly observe the surface brightness (SB); i.e., the X-ray emissivity integrated along the line of sight that has contributions from different spherical shells.
Correlating SB fluctuations with velocity fluctuations provides a way to constrain fluid motions in the ICM. This is a promising approach in absence of direct turbulent velocity 
measurements from high resolution X-ray spectra.

We define the surface brightness as
\begin{equation}
SB(x,y)=\int_{-L/2}^{L/2}n^2(x,y,z)\Lambda[T(x,y,z)]\text{d}z. 
\end{equation}
Note that before performing these calculations, we manually set the density fluctuations from the 
mean values to decay slowly to zero outside a sphere centered at the origin, with a scale radius $L/5$. This is done to impose a realistic spherical symmetry, but its effects are 
moderate and only at the lowest wavenumbers. The procedure is described in detail in appendix \ref{sec:appdx_surf_brightness}.

\subsubsection{Dependence on Mach number}
\label{subsubsec:SB_mach}

\begin{figure}
	\includegraphics[width=\columnwidth]{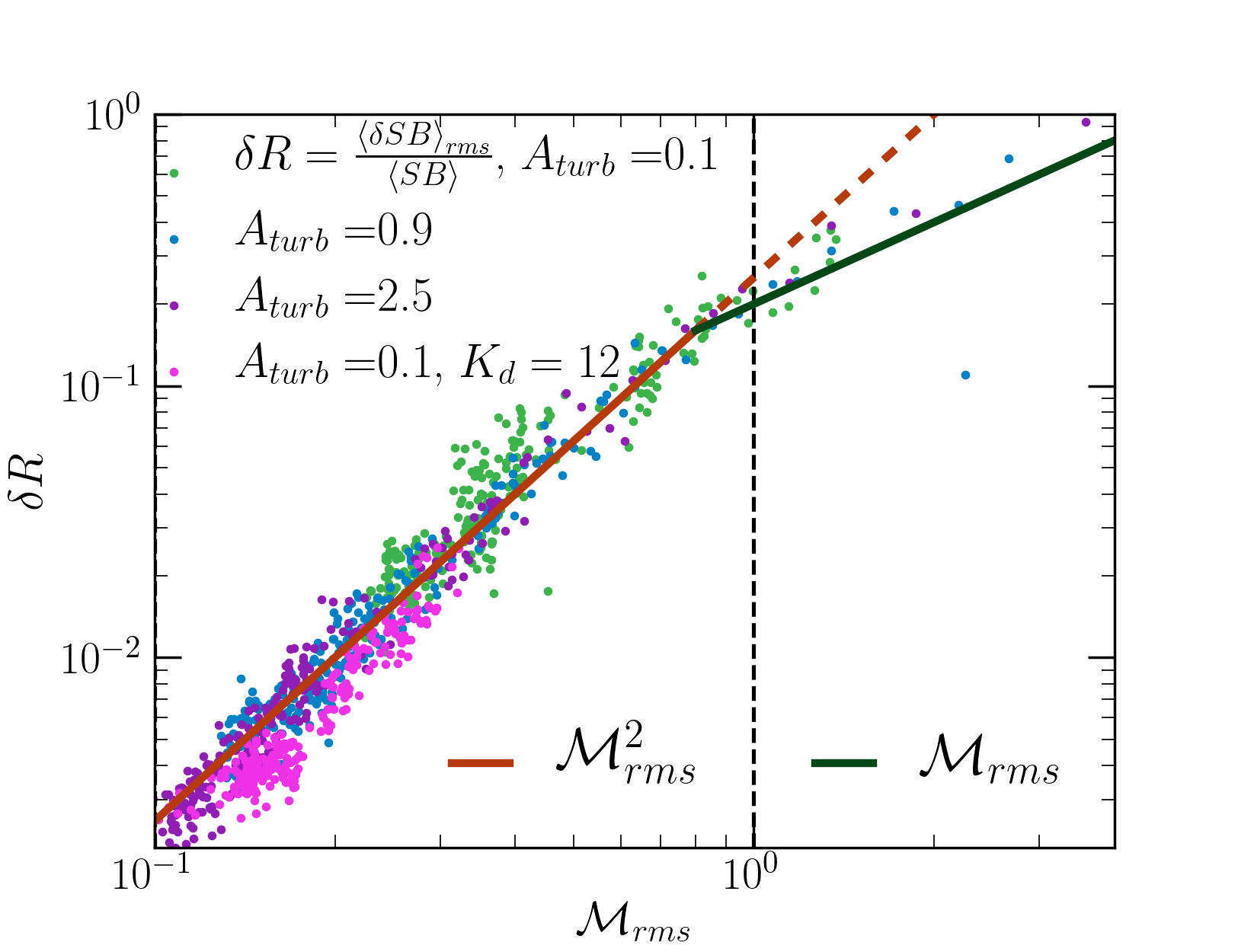}
    \caption[Surface brightness fluctuations vs Mach number]{Surface brightness fluctuations as a function of the rms Mach number for the same runs as \autoref{fig:rho-mach} 
    (these are run for longer times to densely cover the range of $\delta R$ and ${\cal M}_{\rm rms}$). 
    The dependence of surface brightness fluctuations on rms Mach number is similar to that of density and pressure fluctuations in \autoref{fig:rho-mach}; $\delta(SB)$/$SB$ varies 
    as  $\mathcal{M}_{\text{rms}}^2$ in the subsonic regime and is flatter in the supersonic regime. The same scaling is expected for the projected pressure 
    fluctuations probed by the thermal Sunyaev-Zeldovich effect due to the hot ICM.}
    \label{fig:sb-mach}
\end{figure}

\autoref{fig:sb-mach} shows that the surface brightness fluctuations $\delta (SB)/\mean{SB}$ have a similar dependence on $\mathcal{M}_{\text{rms}}$ as $\delta \rho/\mean{\rho}$. 
In the subsonic regime surface brightness fluctuation amplitude varies as $\mathcal{M}_{\text{rms}}^2$, and in the supersonic regime it is flatter. 

\subsubsection{Surface brightness power spectra}
\label{subsubsec:sb_powerspectra}

\begin{figure}
	\includegraphics[width=\columnwidth]{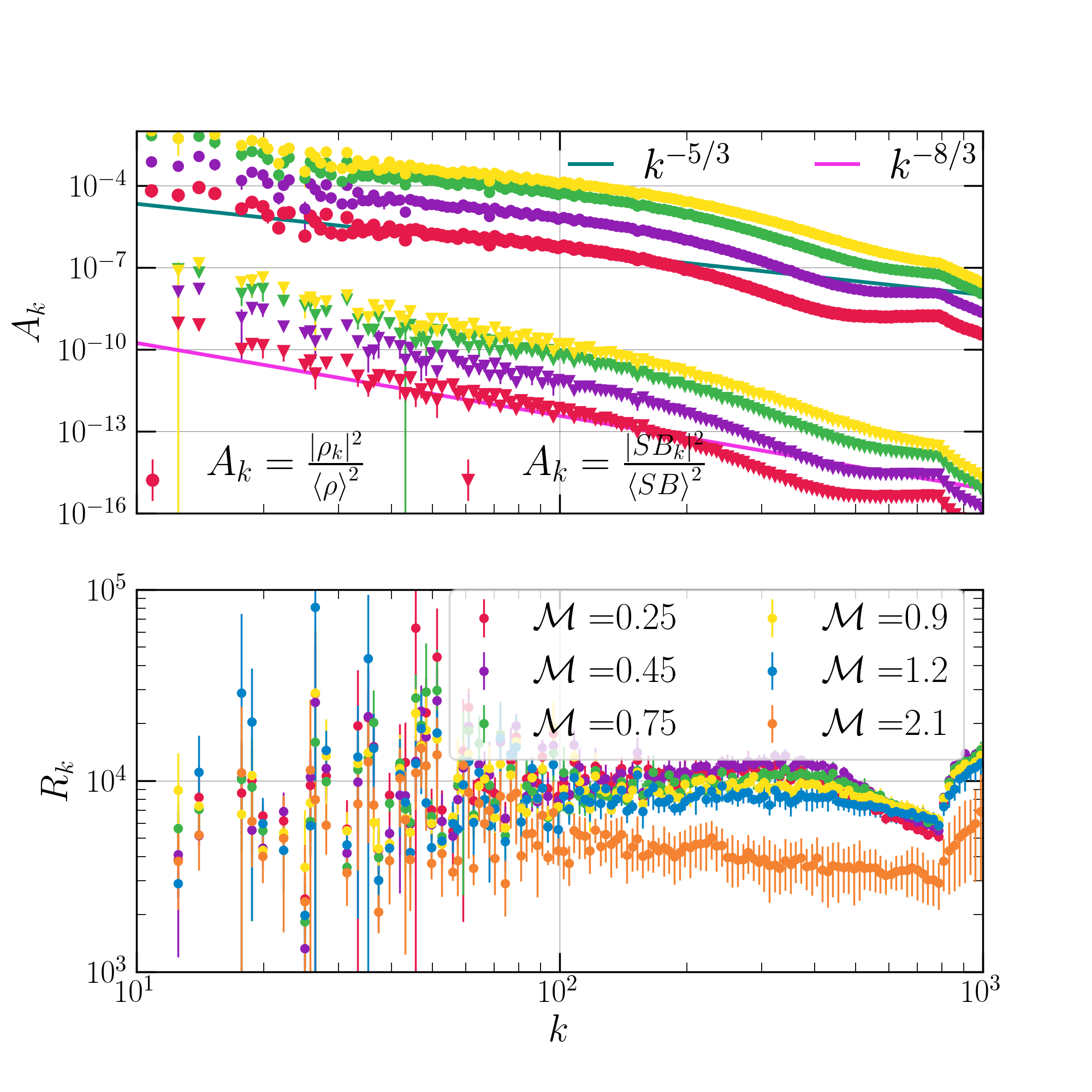}
    \caption[Density and surface brightness power spectra]{ Upper panel: Power spectra of density and surface brightness fluctuations for different $\mathcal{M}_{\text{rms}}$. 
    Density spectra follow the K41 $k^{-5/3}$ scaling in the inertial range and SB power spectrum is steeper by unity. Lower panel: Ratio between 
    density and compensated SB spectra $R_k=\left(\frac{|\rho_k|^2}{\left<\rho\right>^2}\right)/\left(k\frac{|SB_k|^2}{\left<SB\right>^2}\right)$; $R_k$ is constant over the inertial range, showing little variation 
    with $\mathcal{M}_{\text{rms}}$ for  $\mathcal{M}_{\text{rms}}<1$. Some of the error bars here are larger than in \autoref{fig:rho-vel-spectra} because the number of $k$-space points within 2-D annuli 
    are smaller than in 3-D shells for the same bin-size in $k$, and we may be dominated by Poisson noise for low-$k$ bins.}
    \label{fig:rho-sb-spectra}
\end{figure}

\autoref{fig:rho-sb-spectra} shows that the surface brightness power spectra follow a $k^{-8/3}$ scaling in the inertial range, which is steeper by unity than the 
density spectrum $\propto k^{-5/3}$. This is because the number of ${\bf k}$-space points grid points within $\Delta k$ is proportional to $4 \pi k^2 \Delta k$ for spherical shells and to 
$2\pi k \Delta k$ for circular annuli.
Since the power spectra differ by a factor of $k$, the spectral amplitudes of surface brightness and density fluctuations would differ by a factor of $k^{1/2}$. 
This result is in line with the 3D and 2D spectral amplitude relations discussed in section 3 of \citet{churazov2012x}. In the subsonic regime the ratio of relative density 
and compensated surface brightness fluctuations $\left(\frac{|\rho_k|^2}{\left<\rho\right>^2}\right)/\left(k\frac{|SB_k|^2}{\left<SB\right>^2}\right)$ is almost a constant in the inertial range.

\section{Results -- heating balancing cooling} 
\label{sec:ResultsHC}

In the simulations described in this section, we are more faithful to cool core thermodynamics and explicitly balance radiative cooling rate with the sum of turbulent and 
thermal heating rates. Observations show that the hot gas is in rough thermal balance. 
The factor $f_{\text{turb}}$ gives the turbulent heating fraction out of the total (thermal+turbulent) heating. 
The gas is uniformly heated by a constant thermal heating rate density $Q=(1-f_{\text{turb}})\mean{\mathcal{L}}$, 
where $\mean{\mathcal{L}}$ is the average radiative cooling rate of the box. In some of our runs, we seed the gas with initial random density perturbations, using the method 
described in \autoref{subsec:Density-perturbations}. \autoref{tab:cool_runs} lists our thermal balance simulations.

\begin{figure}
	\includegraphics[width=\columnwidth]{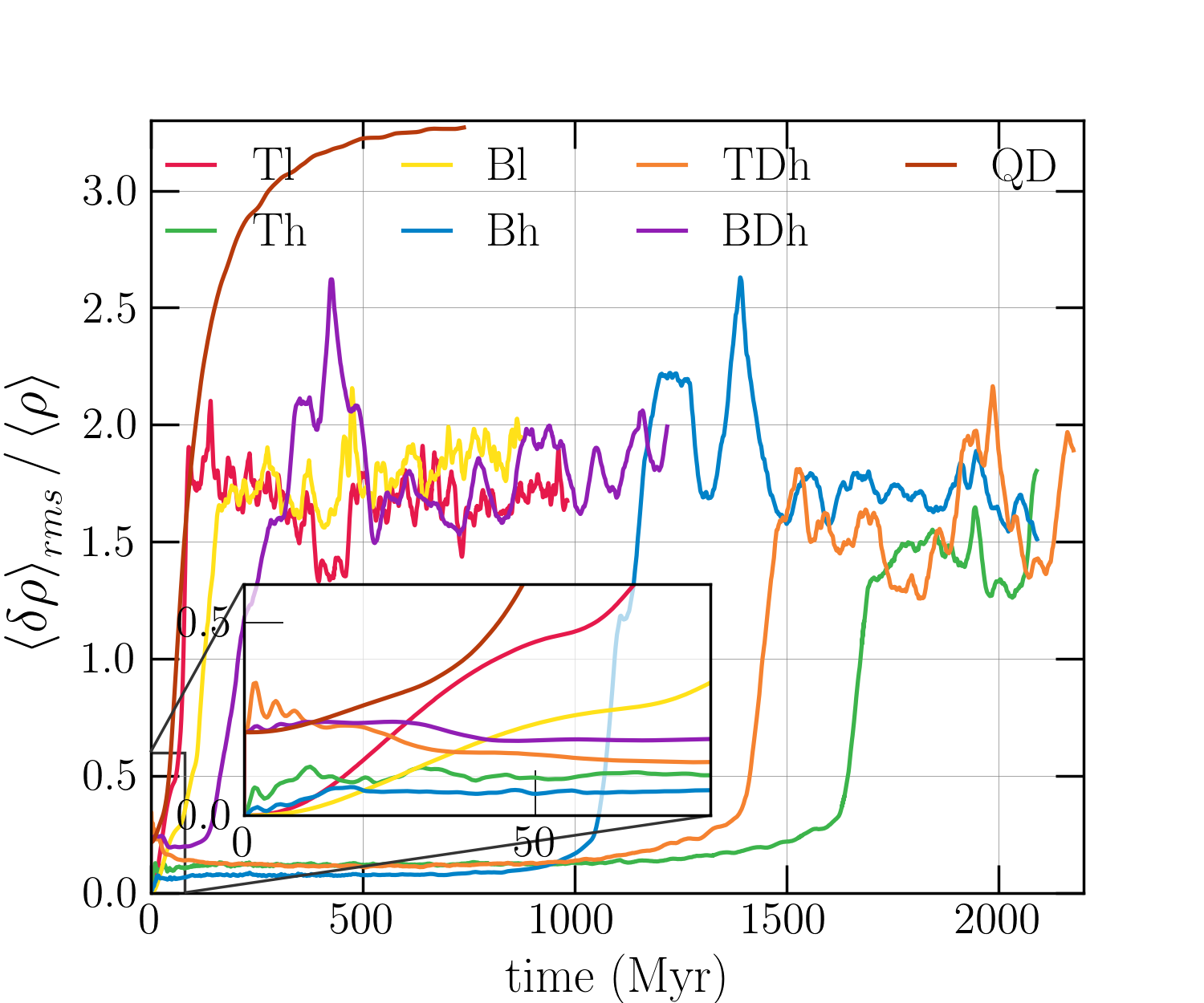}
    \caption{Time evolution of volume-averaged rms density fluctuations for different thermal balance runs. The small flat region at initial times corresponds to the turbulent steady state 
    (seen clearly for the runs that show multiphase gas at later times), and the sharp increase corresponds to multiphase condensation due to thermal instability. Cold gas condenses 
    out at different times for different runs. The inset shows the early time evolution in more detail.}
    \label{fig:del-rho-time}
\end{figure}

\autoref{fig:del-rho-time} shows the time evolution of the volume-averaged rms density fluctuations (normalized to the mean density) in our thermal balance runs. 
Most of these runs show two stages of evolution -- the first being a turbulent 
steady state and the second reflecting thermal instability that leads to multiphase condensation. The first stage occurs after an eddy turnover time scale  
for most of our runs. It depends on the amplitude of forcing, and thus on the parameter $f_{\text{turb}}$ (the fraction of turbulent heating). The second stage of evolution has 
much higher density fluctuations ($\mean{\delta\rho}_{\text{rms}}/\mean{\rho} \geq 1)$. In this stage, the gas separates into hot and cold phases due to thermal instability. 
The multiphase gas formation time scale $(t_{\text{mp}})$ is very different for different parameter choices.

\begin{table*}
	\centering
	\caption{Thermal balance runs}
	\label{tab:cool_runs}
	\resizebox{\textwidth}{!}{
		\begin{tabular}{lcccccccc} 
			\hline
			Label & Resolution & Forcing Amplitude & $K_{\text{driving}}$ &  $f_{\text{turb}}$ & initial $\frac{\mean{\delta\rho}_{\text{rms}}}{\mean{\rho}}$ & $t_{\text{mp}}$(Myrs) & $\sigma_{v_{\text{LOS}}}$(km/s) & Remarks\\
			\hline
			Tl  & $256^3$ & autoscaled & $0<K_{\text{driving}}\leq\sqrt{2}$ & 1.0 & off & 90 & 380 & supersonic gas\\
			Th  & $256^3$ & autoscaled & $K_{\text{driving}}= 12$ & 1.0 & off & 1700 & 255 & long $t_{\text{mp}}$\\
			\hline
			Tlr  & $512^3$ & autoscaled & $0<K_{\text{driving}}\leq\sqrt{2}$ & 1.0 & off & 90 & -- & results converge with Tl\\
			Thr  & $512^3$ & autoscaled & $K_{\text{driving}}= 12$ & 1.0 & off & 1500 & -- & shorter $t_{\text{mp}}$ as compared to Th\\
			\hline
			Bl  & $256^3$ & autoscaled & $0<K_{\text{driving}}\leq\sqrt{2}$ & $0.5$ & off & 160 & 361 & supersonic gas\\
			Bh  & $256^3$ & autoscaled & $K_{\text{driving}}= 12$ & $0.5$ & off & 1200 & 228 & long $t_{\text{mp}}$\\
			\hline
			QD  & $256^3$ & 0 & n/a & $0$ & $0.2$ & 40 & 32 & immobile cold gas clumps\\
			TDh  & $256^3$ & autoscaled & $K_{\text{driving}}=12$ & 1.0 & $0.2$ & 1500 & 260 & long $t_{\text{mp}}$\\
			BDh  & $256^3$ & autoscaled & $K_{\text{driving}}=12$ & $0.5$ & $0.2$ & 400 & 226 & subsonic gas, reasonable $t_{\text{mp}}$\\
			\hline
			BDh2  & $256^3$ & autoscaled & $K_{\text{driving}}=12$ & $0.1$ & $0.2$ & 240 & 165 & reproduces Hitomi velocity profile\\
			BDh3  & $256^3$ & autoscaled & $K_{\text{driving}}=12$ & $0.3$ & $0.2$ & 320 & 202 & subsonic gas, reasonable $t_{\text{mp}}$\\
			BDh4  & $256^3$ & autoscaled & $K_{\text{driving}}=12$ & $0.7$ & $0.2$ & 500 & 254 &subsonic gas, reasonable $t_{\text{mp}}$\\
			BDh5  & $256^3$ & autoscaled & $K_{\text{driving}}=12$ & $0.9$ & $0.2$ & 700 & 264 & long $t_{\text{mp}}$\\
			\hline
	\end{tabular}}
	\justifying \\ \begin{footnotesize} In the labels, T stands for pure turbulent heating ($f_{\text{turb}}=1$), Q denotes pure thermal heating ($f_{\text{turb}}=0$) and B stands for both thermal and turbulent heating, r denotes a resolution of $512^3$ grid points (all other runs use a grid with $256^3$ grid points), l denotes driving at low $k$s ($0< K_{\text{driving}}\leq\sqrt{2}$), h denotes driving at high $k$s ($K_{\text{driving}}= 12$), D denotes initial density perturbations with $|\rho_k|=Ak^{-1/3}$ ($\sqrt{2}\leq k\leq 12$,  $A$ is a constant amplitude). $\sigma_{v_{\text{LOS}}}$ represents the velocity dispersion along the line of sight.
	\end{footnotesize} 
	
\end{table*}

\begin{figure}
	\centering
	\includegraphics[width=\columnwidth]{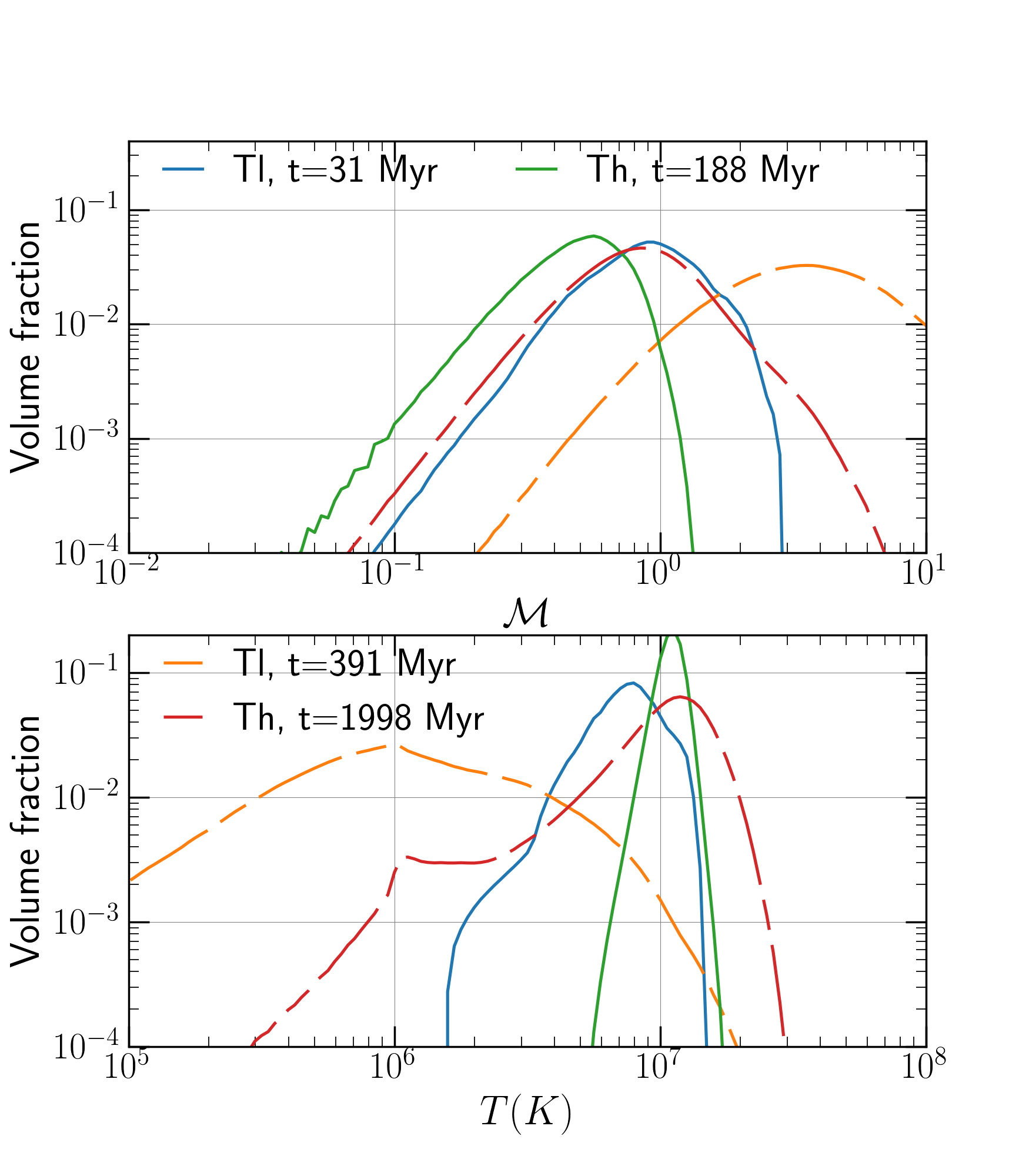}
	\caption[$\mathcal{M}$ and T distribution for turbulent heating]{Volume probability distribution function (PDF) of Mach number ($v/c_s$; upper panel) and temperature (lower panel) in the turbulent steady 
	state (before condensation) and after multiphase gas formation for pure turbulent driving runs: high $K_{\text{driving}}$ (Th) and low $K_{\text{driving}}$ (Tl). Note that the amount of gas at 
	intermediate temperatures and the spread of the PDFs are different for different runs/times. At late times we see a narrow peak (corresponding to the hot phase) and a slight bump 
	(for gas at $T_{\rm cutoff}$) in the Mach number distribution for Th, whereas a single broad peak at ${\cal M}\sim 3$ is observed for Tl.}
	\label{fig:mach-temp-dist-T}
\end{figure}

\begin{figure}
	\subfloat[]{ 
		\includegraphics[width=0.45\columnwidth]{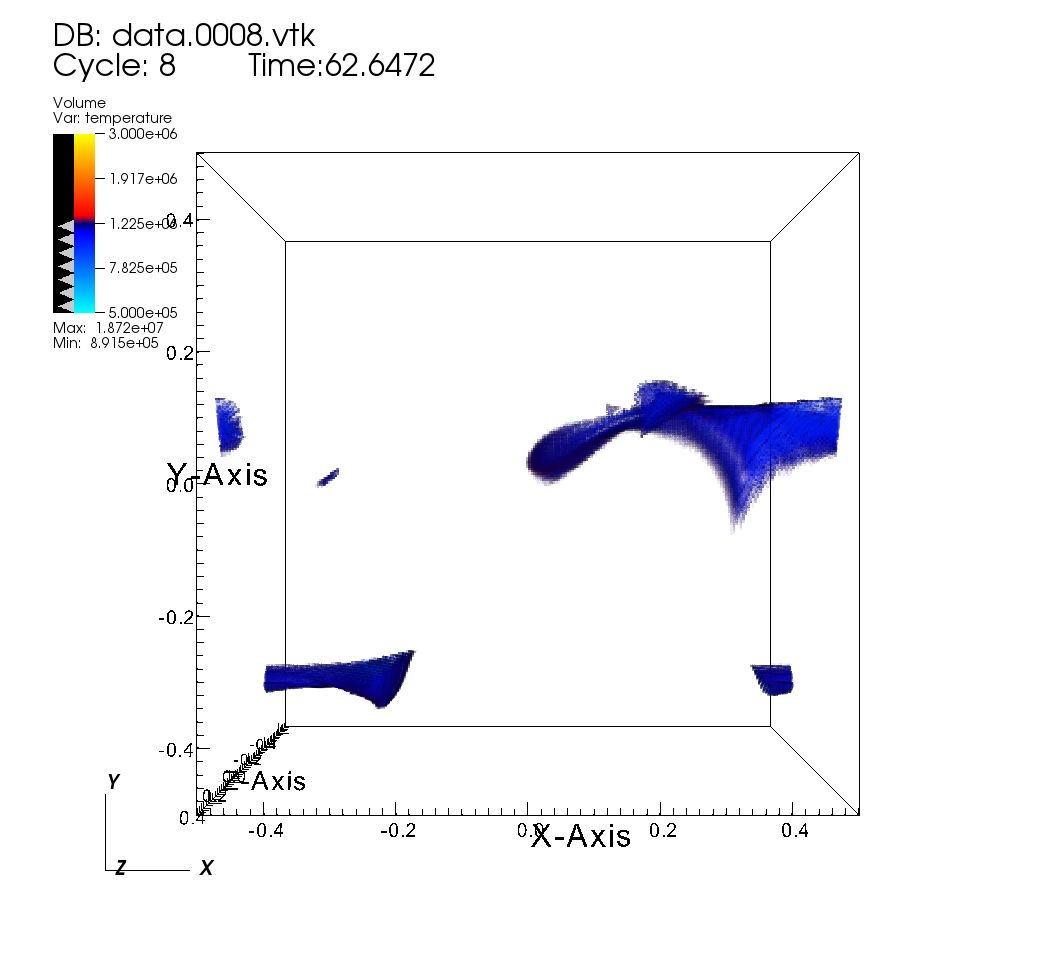} 
	} 
	\hfill 
	\subfloat[]{ 
		\includegraphics[width=0.45\columnwidth]{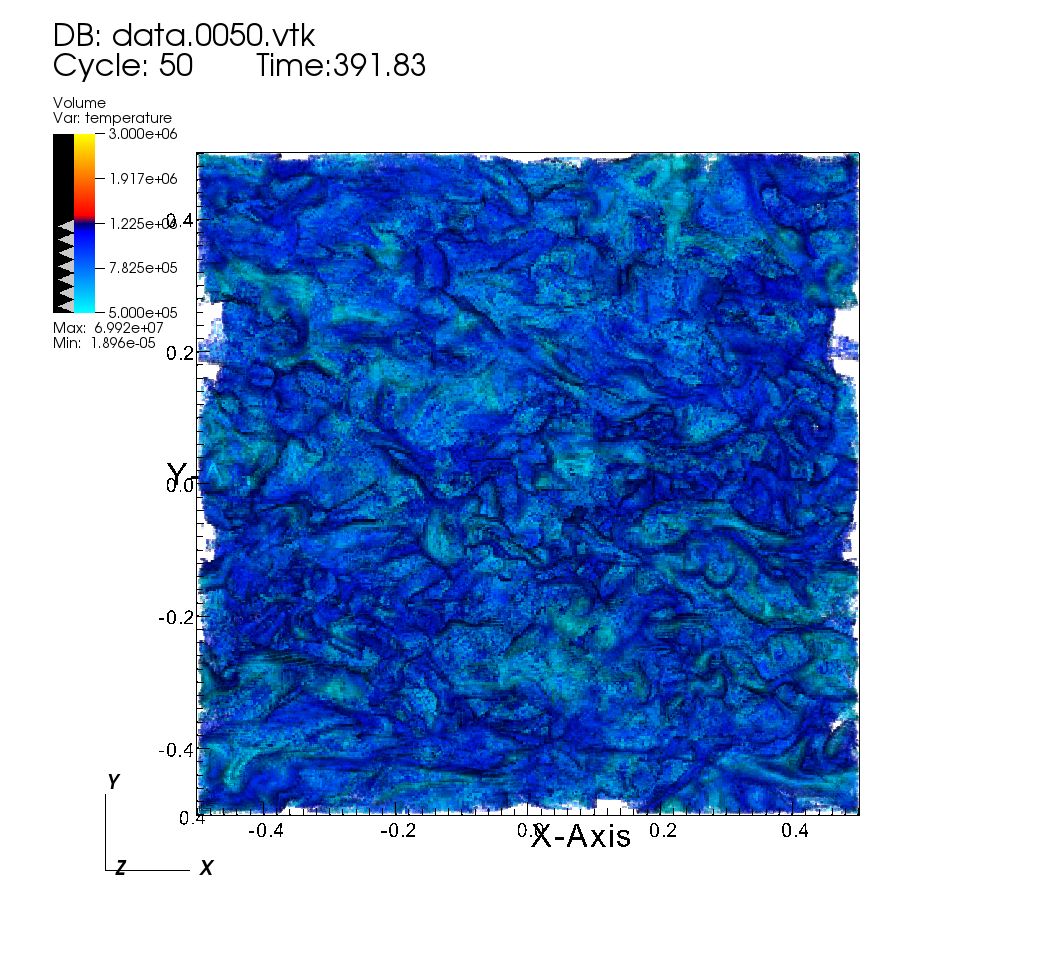} 
	} 
	\hfill 
	\subfloat[]{ 
		\includegraphics[width=0.45\columnwidth]{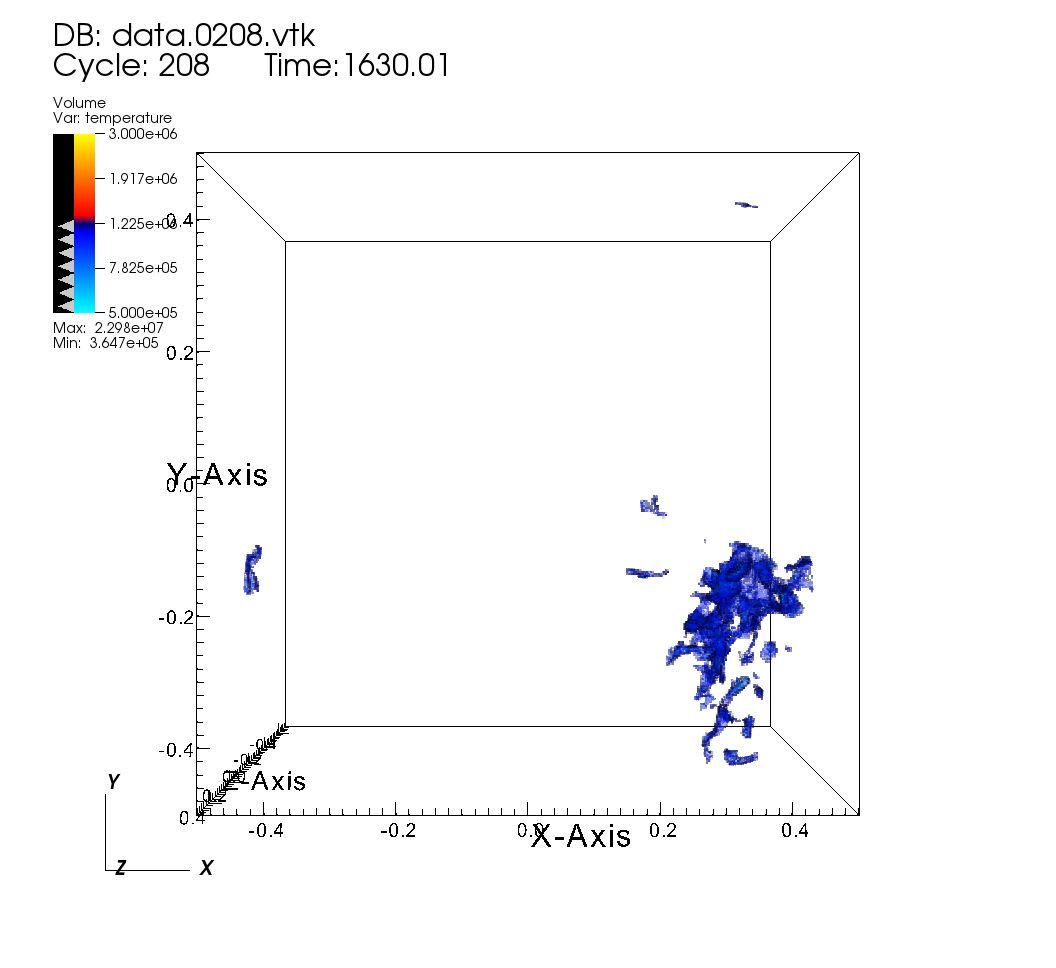} 
	} 
	\hfill 
	\subfloat[]{ 
		\includegraphics[width=0.45\columnwidth]{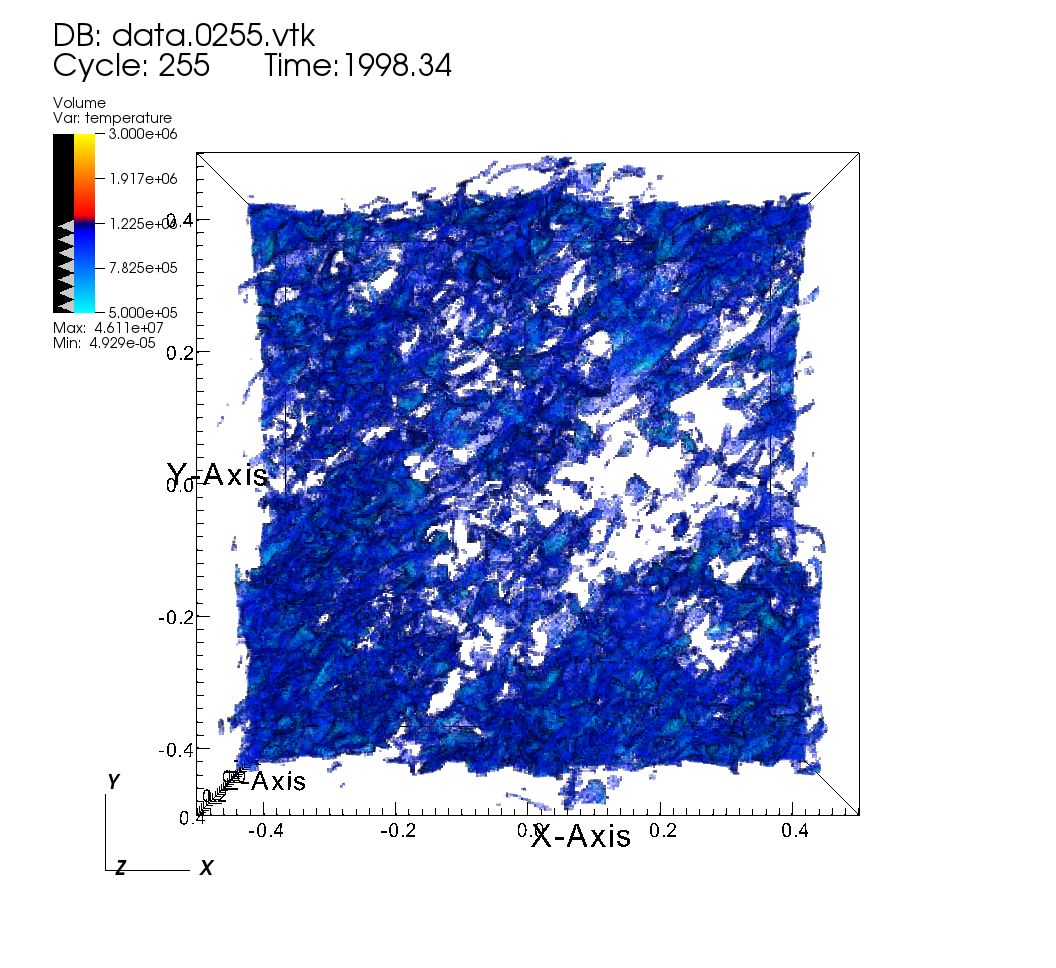} 
	} 
	\hfill 
	\caption[Cold gas volume rendering for pure turbulent heating]{Volume rendering of cold gas for Tl and Th runs. Gas having temperature greater than $1.22\times10^6K$ is set to be transparent, 
	so that we show only the cold gas. The upper panels correspond to large scale driving (Tl), and the lower panels to small scale driving (Th). The left panels show gas just after cold gas 
	starts condensing (62.6, 1630.0 Myr) and the right panels show cold gas at a later time (391.8, 1998.3 Myr). Turbulence plays the dual role of seeding density perturbations and 
	mixing density/temperature 
	inhomogeneities.}\label{fig:T-l-h}
\end{figure}

\subsection{Pure turbulent heating (Tl \& Th)}
\label{subsec:T}
For runs Tl (low-$k$ driving) and Th (high-$k$ driving; see \autoref{tab:cool_runs}), we use $f_{\text{turb}}=1$ (i.e., 
turbulent heating fully compensates energy 
losses due to radiative cooling at each time step). We do not initialize density perturbations in these runs (they are seeded by turbulence itself). Both runs 
with driving at large and small length scales show multiphase gas. While the run with large driving scales (Tl) has $t_{\text{mp}}\approx 80$ 
Myr, for small scale driving (Th) $t_{\text{mp}}\approx 1700$ Myr ($\sim$20 times longer!). The time $t_{\text{mp}}$ can be directly measured from the plot of 
rms density perturbation versus time (\autoref{fig:del-rho-time}), which grows by an order of magnitude when  
multiphase gas condenses. Local thermal instability can lead to cold gas condensation and nonlinear density perturbations, 
but it may take several cooling times.  

\autoref{fig:mach-temp-dist-T} shows the Mach number and temperature distributions for the two runs Tl and Th before and after multiphase condensation. Compared to 
Th, Tl has a fair amount of gas at intermediate temperatures (i.e., between $T_{\text{hot}}$ and $T_{\text{cutoff}}$, where $T_{\text{hot}}$ is the temperature 
of the hot phase). Also, $T_{\text{hot}}$ is smaller for Tl as compared to Th. The cold peak
is more prominent for large scale driving (Tl). For Tl most of the gas is supersonic at time $t > t_{\text{mp}}$, with peak at ${\cal M} \sim 3$. For small scale driving (Th), 
the peak in Mach number distribution is at ${\cal M} \sim 1$, with a small bump at ${\cal M} \sim 3$.

Turbulence mixes up gas at all length scales starting from the driving scale. Hence, large length scale driving mixes up the gas better on larger scales than small length scale driving. 
By mixing, turbulence smoothens the temperature PDF which is driven towards bimodality due to thermal instability. In these runs, turbulent driving itself generates larger amplitude 
of density fluctuations, since in the inertial range 
$\delta \rho_l \propto l^{1/3}$. Denser regions have faster runaway cooling, which leads to quick formation of multiphase gas, at around $t=80$Myr.  Top right panel of \autoref{fig:T-l-h} shows that the 
distribution of cold gas in run Tl is rather uniform throughout the simulation domain.

For small scale driving, it takes much longer than the cooling time scales for multiphase gas condensation $(\approx 1700~\text{Myr}$; \autoref{fig:del-rho-time}). 
In this case, cold gas condenses in more localised regions, and the cloud grows around it. 
In \autoref{fig:mach-temp-dist-T} for Th the narrower transonic peak corresponds to hot gas and a small supersonic bump $(\mathcal{M}\approx 3)$ to cold gas clouds.
We can attribute the long $t_{\rm mp}$ (time for multiphase condensation) in the small scale driving run (Th) 
 to small density perturbations generated by small scale driving $(\delta \rho_l\propto l^{1/3})$ (e.g., see the inset in \autoref{fig:del-rho-time}; c.f. \autoref{fig:rho-prs-mach-cool}). 
 These small density perturbations are quickly mixed up by turbulence itself before runaway cooling can happen, thus preventing the formation of larger cool and overdense regions. 
 Later in sections \ref{subsec:B} \& \ref{subsec:D} we
 show that cold gas condenses early if large density perturbations and thermal (non-turbulent) heating are present. 

In cool cluster cores the gas temperature distribution is bimodal (in reality the cooler phase will be emitting in H$\alpha$ and CO and not in X-rays), 
and observations show that the hot ICM is subsonic (\citealt{aharonian2016quiescent}). 
From our simulations, we conclude that it is unlikely that pure turbulent driving on cluster core length scales (10s of kpc) can balance radiative losses in 
the core for $\sim 1$ keV clusters,
since this scenario gives a large amount of gas at intermediate temperatures and supersonic turbulence in the hot phase (subject to our assumptions 
as listed in \autoref{subsec:comparing_results}).
Turbulent driving at small length scales could be important, except that in these runs multiphase gas takes too long to condense out and the Mach number peak
in the hot phase is still larger than observations. 
In the following subsection (\autoref{subsec:B}), 
we look at the impact of introducing uniform thermal (non-turbulent) heating on these simulations. 

\begin{figure}
	\includegraphics[width=\columnwidth]{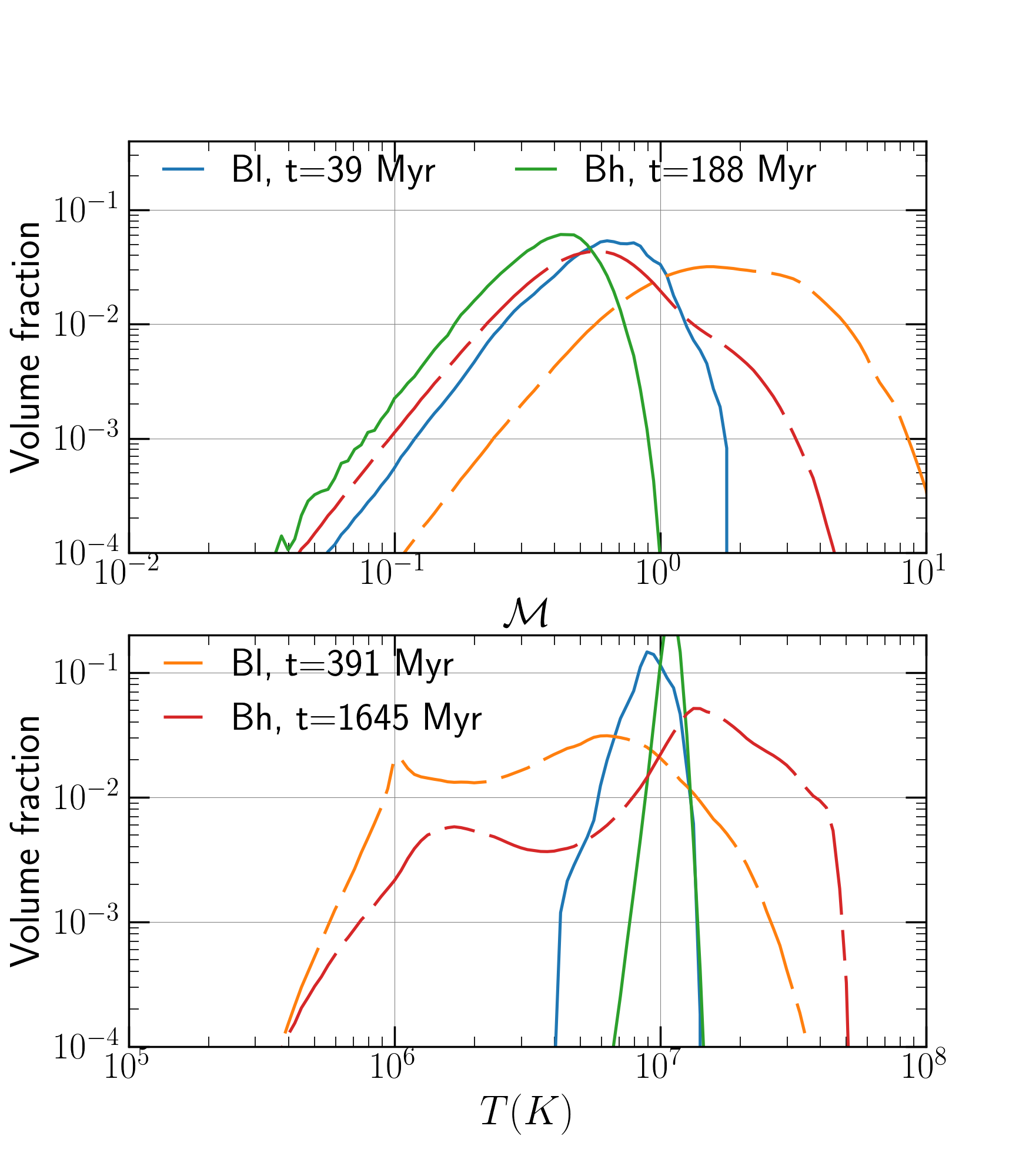}
	\caption{Mach number ($v/c_s$; upper panel) and temperature (lower panel) PDFs in the turbulent steady state and after multiphase condensation for runs with both thermal and 
	turbulent heating $(f_{\text{turb}}=0.5)$ and driving at high (Bh) and low (Bl) $k$s. These runs show more bimodality in temperature distribution compared to their pure turbulent 
	heating counterparts in \autoref{fig:mach-temp-dist-T}.}
	\label{fig:mach-temp-dist-B}
\end{figure}

\begin{figure}
	\subfloat[]{ 
		\includegraphics[width=0.45\columnwidth]{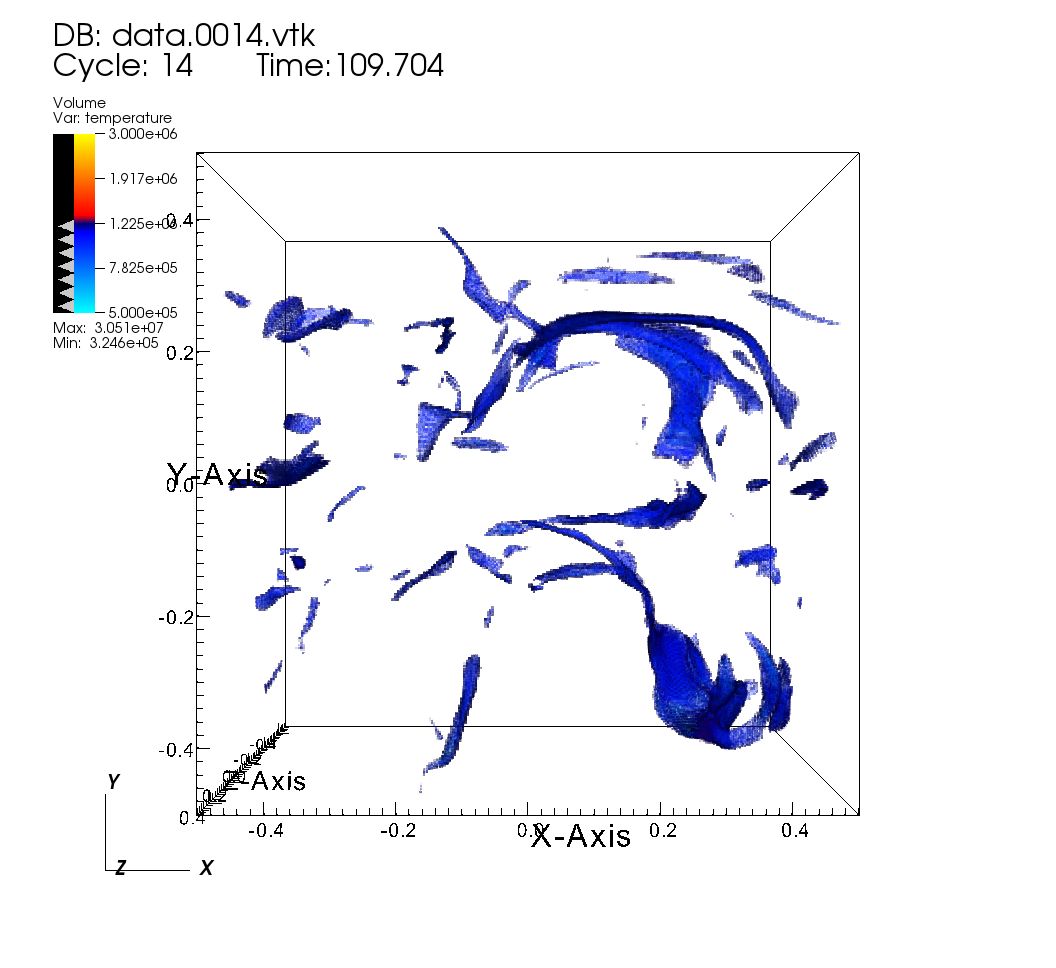} 
	} 
	\hfill  
	\subfloat[]{ 
		\includegraphics[width=0.45\columnwidth]{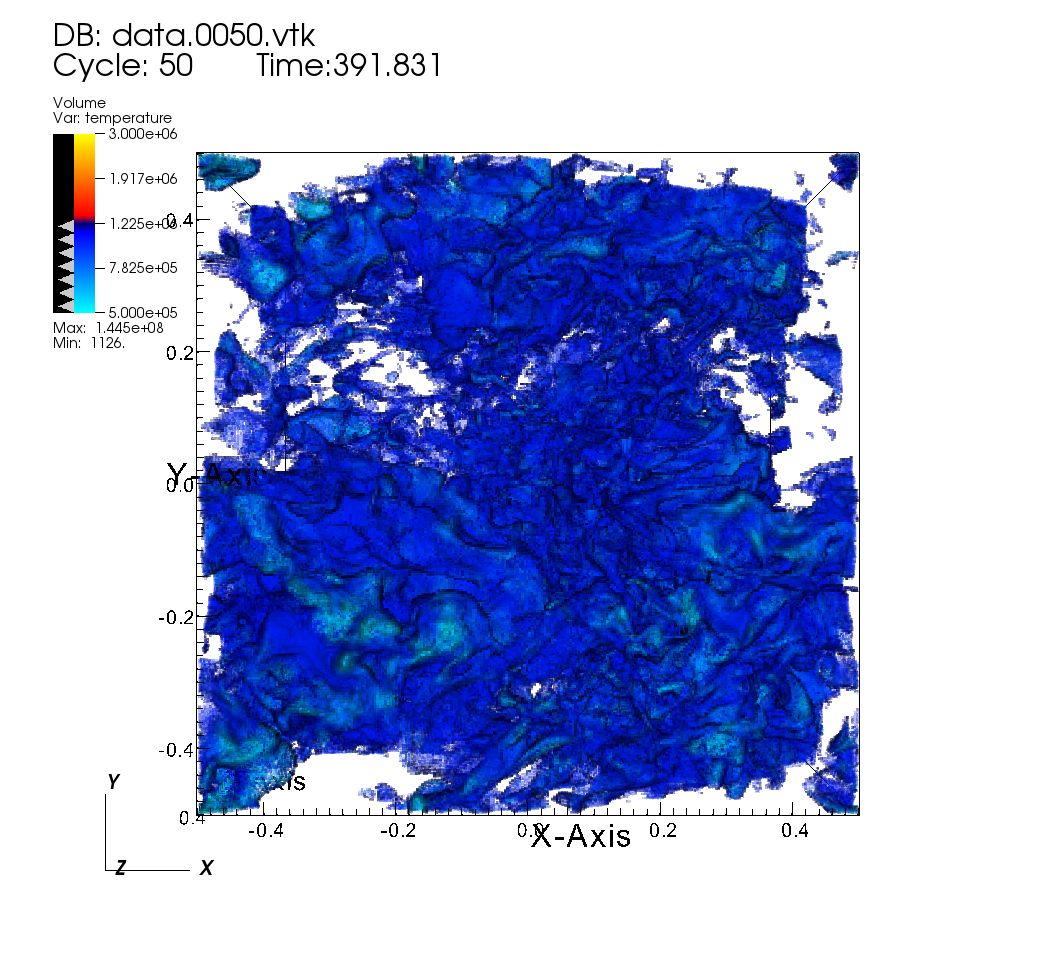} 
	} 
	\hfill 
	\subfloat[]{ 
		\includegraphics[width=0.45\columnwidth]{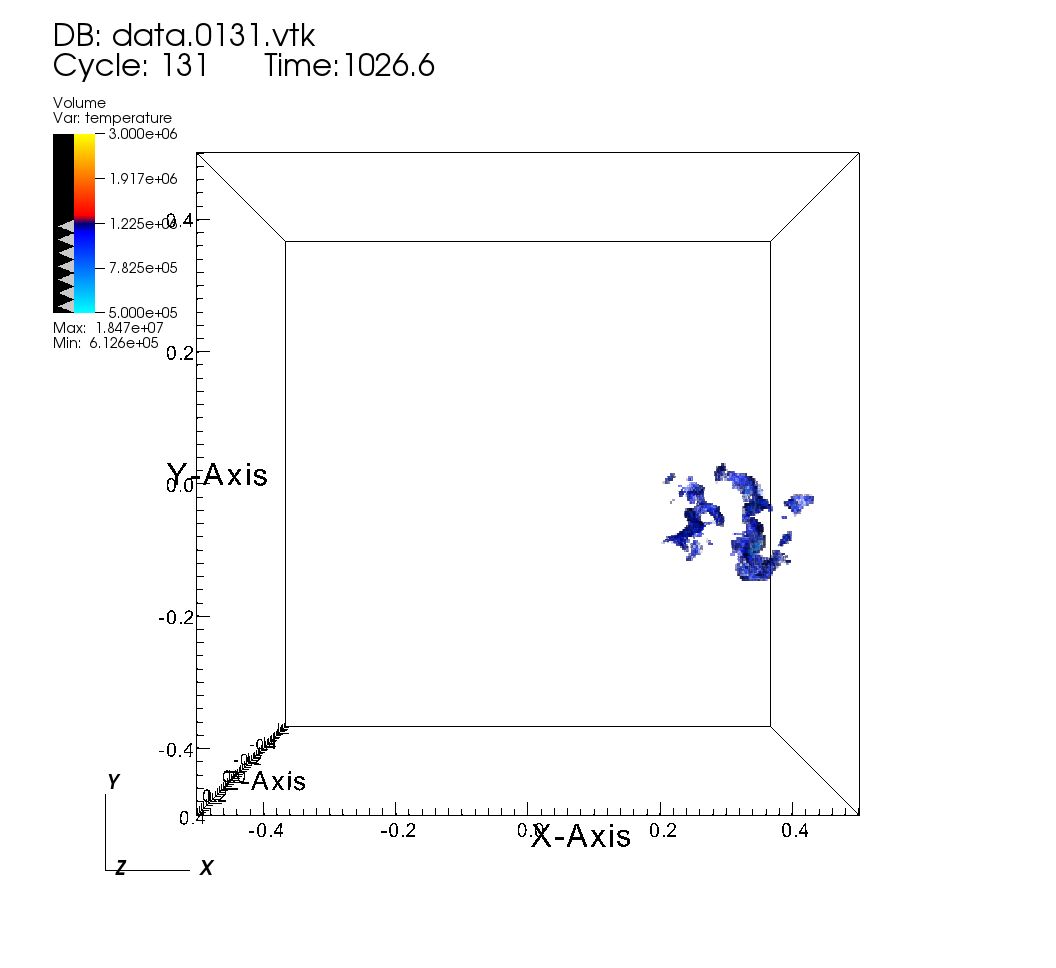} 
	} 
	\hfill
	\subfloat[]{ 
		\includegraphics[width=0.45\columnwidth]{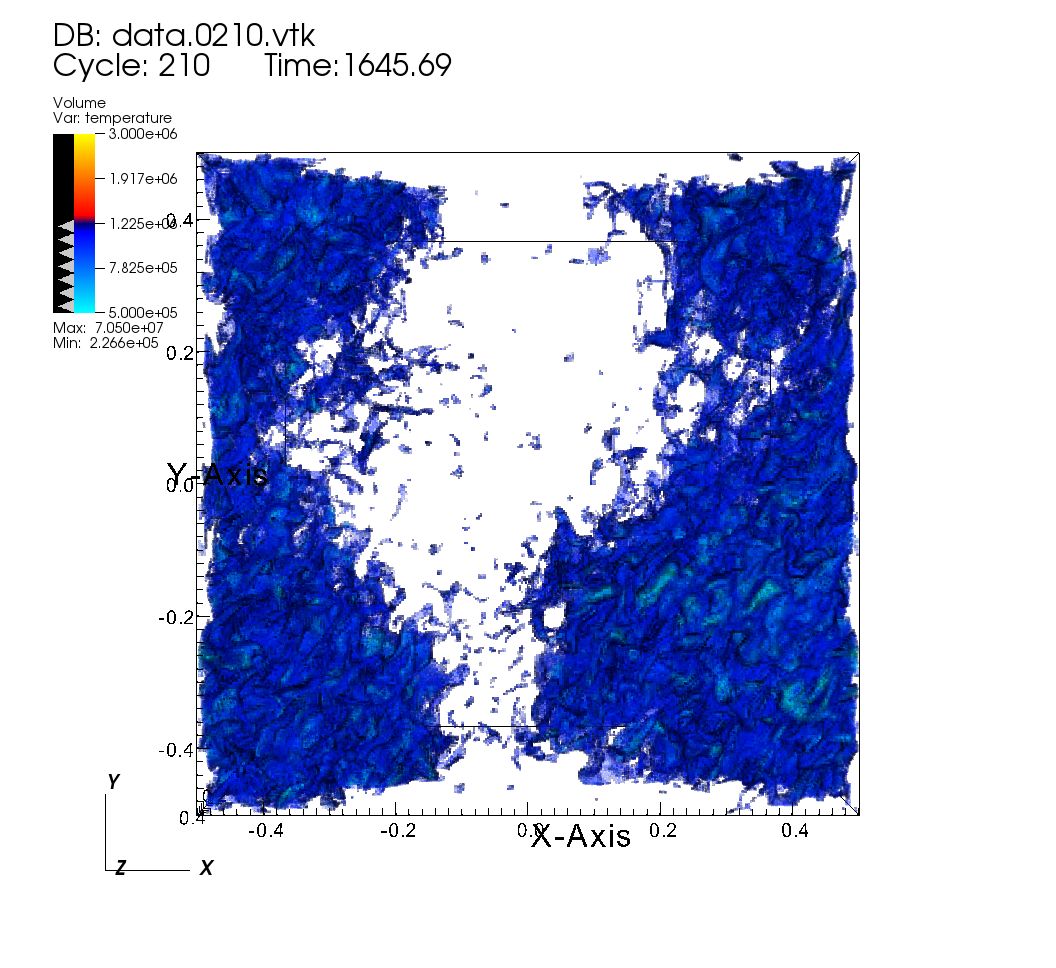} 
	} 
	\hfill 
	\caption[Cold gas volume rendering for both thermal and turbulent heating]{Volume rendering of cold gas ($T<1.22\times10^6K$) for runs with equal turbulent and thermal heating (Bl and Bh). 
	The upper panels correspond to large scale driving (Bl), and the lower panels to small scale driving (Bh). The left panels show gas just after cold gas starts condensing, and the right panels show 
	cold gas at a later time. These are qualitatively similar to the corresponding Tl and Th plots in \autoref{fig:T-l-h}.}\label{fig:B-l-h}
\end{figure}

\subsection{Both thermal and turbulent heating (Bl \& Bh)} \label{subsec:B}
For the runs Bl and Bh we use $f_{\text{turb}}=0.5$; i.e., half of the cooling losses are balanced by turbulent heating and the other half by the heat added uniformly throughout the volume. 

\autoref{fig:del-rho-time} shows that for large scale driving cold gas condenses out early, at around 160 Myr. \autoref{fig:mach-temp-dist-B}
shows that the amount of gas having temperature below the cut-off temperature is lower than the corresponding pure turbulent heating runs (shown in Fig. \ref{fig:mach-temp-dist-T}), 
because of a smaller turbulent forcing. However, we still have a lot of gas at intermediate temperatures, and a broad supersonic peak in the Mach number distribution.

For smaller scale forcing (high $K_{\text{driving}}$), cold gas forms a bit earlier (at around 1200 Myr, compared to 1700 Myr for pure turbulent heating runs). This time is still an order of magnitude longer
than the cooling time. 
For Bh runs (as compared to Th) cooler regions get more time to grow before they are mixed up with hotter regions, which leads to large density fluctuations and smoother temporal evolution. 
\autoref{fig:mach-temp-dist-B} shows that the hot gas is fairly subsonic $(\mathcal{M}\approx 0.6)$, and the cold gas is modestly supersonic $(\mathcal{M}\approx 2.5)$ for these simulations. 
The distribution of gas in different phases is more bimodal, with less gas at intermediate temperatures, as compared to runs with pure turbulent heating (compare Figs. \ref{fig:mach-temp-dist-T} 
\& \ref{fig:mach-temp-dist-B}). The volume rendering plots in \autoref{fig:B-l-h} are similar in nature to those of pure turbulent driving in \autoref{fig:T-l-h}.

From the results of these simulations, we conclude that other thermal heating mechanisms that do not drive strong turbulence (e.g., thermal conduction [e.g., \citealt{wagh2014}], 
turbulent mixing [e.g., \citealt{hillel2017}], cosmic ray streaming [e.g., \citealt{guo2008}], shocks/sound waves [e.g., \citealt{ruszkowski2004}]) 
play an important role in closing the AGN feedback loop (at least in an average sense). Non-turbulent heating  leads to more bimodality in temperature distribution and subsonic gas velocities
 in the hot phase (for small scale driving runs). In the next subsection, we introduce initial density perturbations (over and above what is produced by turbulence), and assess their impact on the 
 multiphase gas.

\subsection{Initial density perturbations (QD, TDh \& BDh)}
\label{subsec:D}

The density perturbations in the ICM may be primarily seeded by sources other than turbulence such as cooling/heating (as in our simulations presented in \autoref{sec:ResultsHC}),
galaxy wakes, rising bubbles and sloshing.
Therefore, for the runs discussed in this section, we initialize isobaric (since sound crossing time over the cluster core scales is shorter than the cooling time) 
density perturbations according to the prescription in \autoref{subsec:Density-perturbations}. 
In this section we discuss the following runs with initial density perturbations: a pure thermal heating run QD ($f_{\text{turb}}=0$, D stands for initial density perturbations), 
a pure turbulent 
run with small-scale driving TDh $(f_{\text{turb}}=1, \text{ }K_{\text{driving}}=12)$, and a thermal$+$turbulent heating run BDh $(f_{\text{turb}}=0.5, \text{ }K_{\text{driving}}=12)$.
We focus on small-scale driving because the Mach number in the hot phase is smaller (and closer to observations) than large-scale driving.

The amplitude of relative initial density perturbations $\mean{\delta\rho}_{\text{rms}}/\mean{\rho}$ is $0.2$, roughly twice the rms density perturbations 
in the turbulent steady state of run Th before cold gas condensation (compare Th and TDh in Fig. \ref{fig:del-rho-time}). We have also tried runs with smaller initial density 
perturbations, which only show a slightly longer $t_{\text{mp}}$, but the Mach number and temperature distributions are similar to the run with smaller/without any density perturbations. 
Thus, small density perturbations do not significantly affect the occurrence of multiphase gas. 

\subsubsection{Thermal heating only (QD)}

This run is similar to the simulations presented in \citet{sharma2010thermal}, in that there is no externally imposed turbulence and the fluid motions are caused by thermal 
instability itself. The key differences are that our simulations are 3-D hydro, while the earlier paper was based on 2-D MHD runs.  \autoref{fig:del-rho-time} shows that cold 
gas starts condensing out at around $t\approx 40$ Myr, comparable to the cooling time. The 
rms perturbations are larger and much smoother in time compared to the runs with turbulence because turbulence mixes the phases in latter, preventing a large
stationary density/temperature contrast.
 
Due to much weaker turbulence in this run, the temperature PDF also shows a strong bimodality in \autoref{fig:mach-temp-dist-D}. 
There is much less gas at intermediate temperatures, 
almost no gas at temperatures below $T_{\text{cutoff}}$, but a tail at large temperatures going as high as $2\times 10^8$ K. All this is a consequence of much weaker turbulence.  
\autoref{fig:mach-temp-dist-D} shows that the initial Mach numbers are very low $(\approx10^{-2})$.
The Mach number PDF even at $t>t_{\text{mp}}$ shows a single broad peak below $\mathcal{M}=0.1$, for both the hot and cold phases. 
The flow is entirely subsonic, including the gas in the cold ($\sim 10^6$ K) phase. 
 
The volume-rendering of cold gas in \autoref{fig:D} (top panels) for the thermal heating run shows that the clouds of cold gas grow at the same location as the initial density peaks. 
The clumps merely grow with time, and have little or no motion, as expected from their low Mach numbers. In the absence of additional driving, the cold and hot gas phases remain well 
separated in space and in density/temperature.

\subsubsection{Small scale driving (TDh)}

\begin{figure}
	\centering
	\includegraphics[width=0.9\columnwidth]{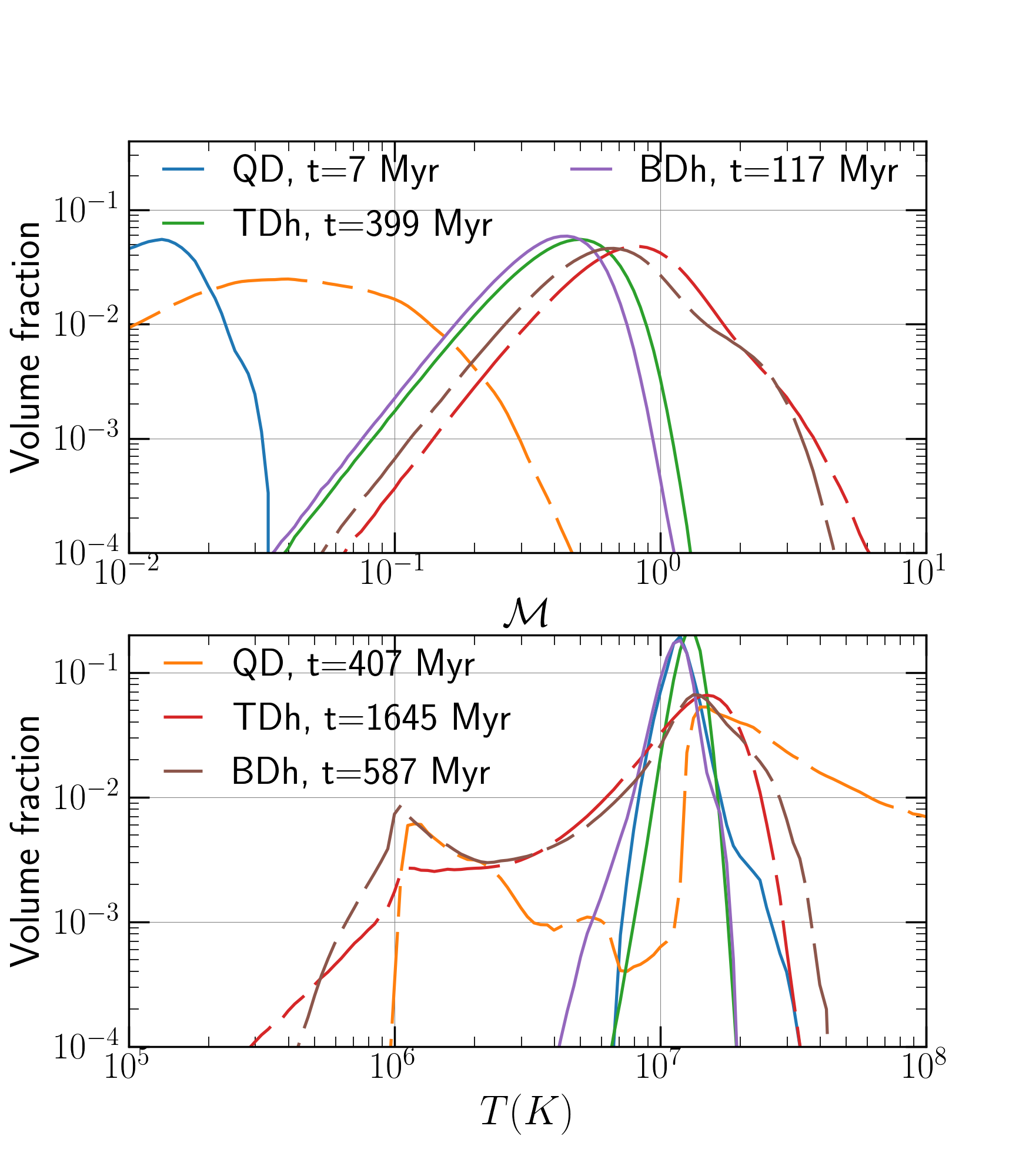}
	\caption[$\mathcal{M}$ and T distribution for initialized density perturbations]{Mach number ($v/c_s$; upper panel) and temperature (lower panel) PDF before and 
	after multiphase gas condensation for runs with initial density perturbations, QD, TDh and BDh. Turbulent forcing for these runs, wherever included, is at small scales. 
	The Mach number for pure thermal heating run (QD) is low, with a single broad peak. This run shows very less gas at intermediate temperatures, and some gas even at 
	$T>10^8 K$. With driven turbulence (TDh, BDh), we have a lot more gas at intermediate temperatures, and at temperatures below $T_{\text{cutoff}}$. The Mach 
	number of these runs is higher, with the cold phase being supersonic. The degree of bimodality (among turbulent forcing runs) is higher for turbulent$+$thermal heating.}
	\label{fig:mach-temp-dist-D}
\end{figure}

\begin{figure}
	\subfloat[]{ 
		\includegraphics[width=0.45\columnwidth]{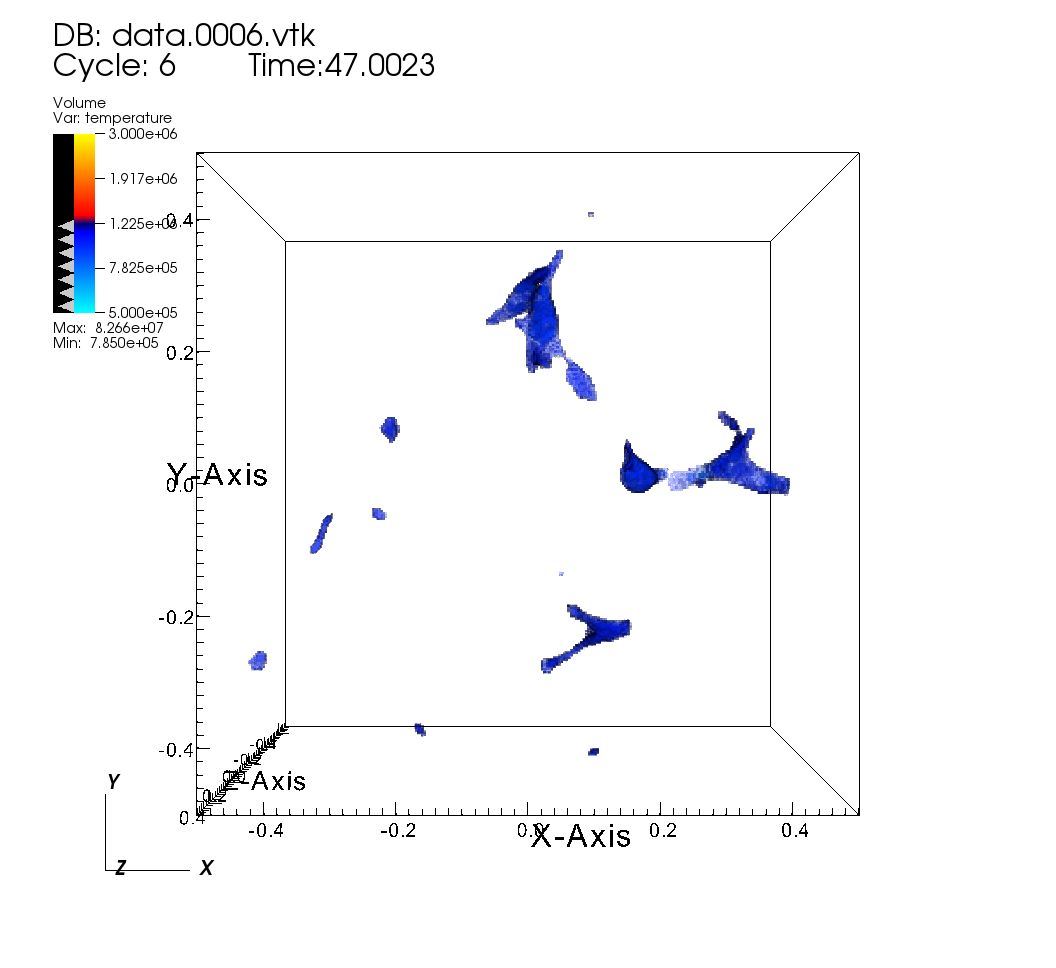} 
	} 
	\hfill  
	\subfloat[]{ 
		\includegraphics[width=0.45\columnwidth]{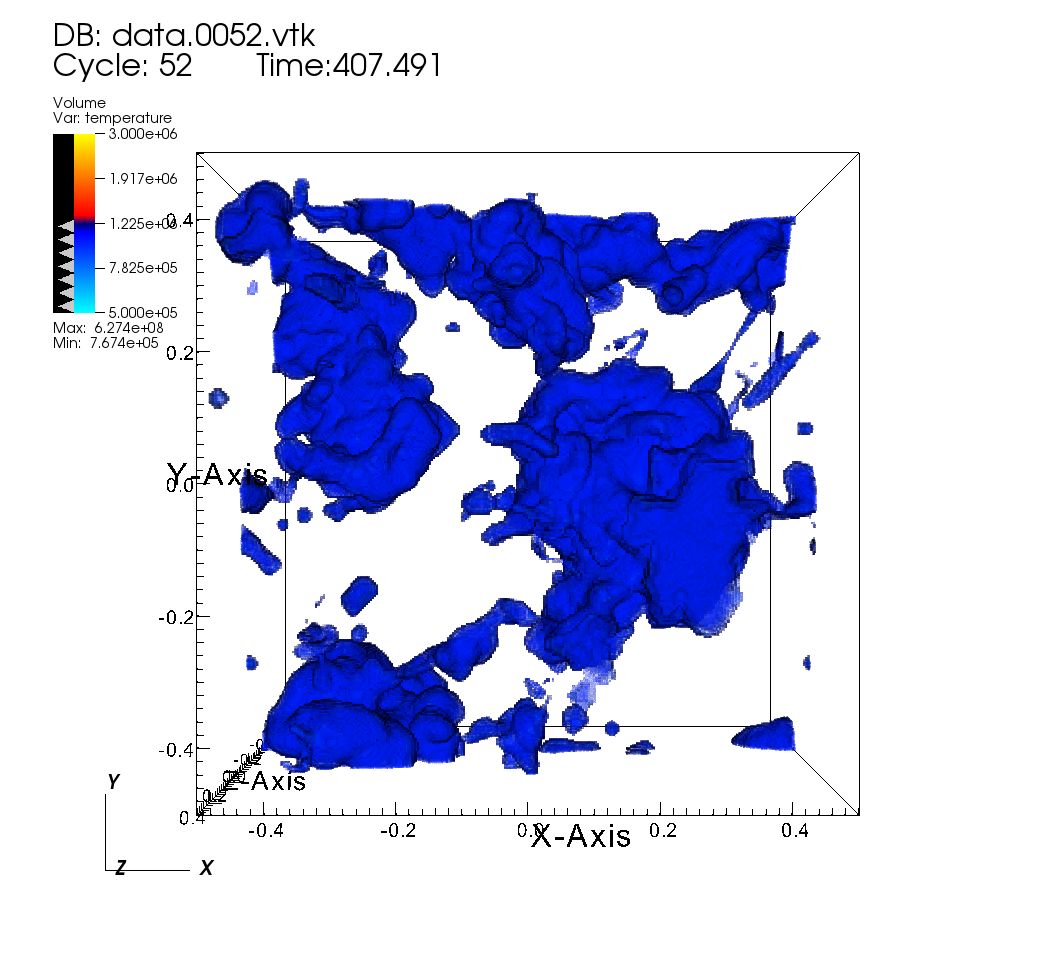} 
	} 
	\hfill 
	\subfloat[]{ 
		\includegraphics[width=0.45\columnwidth]{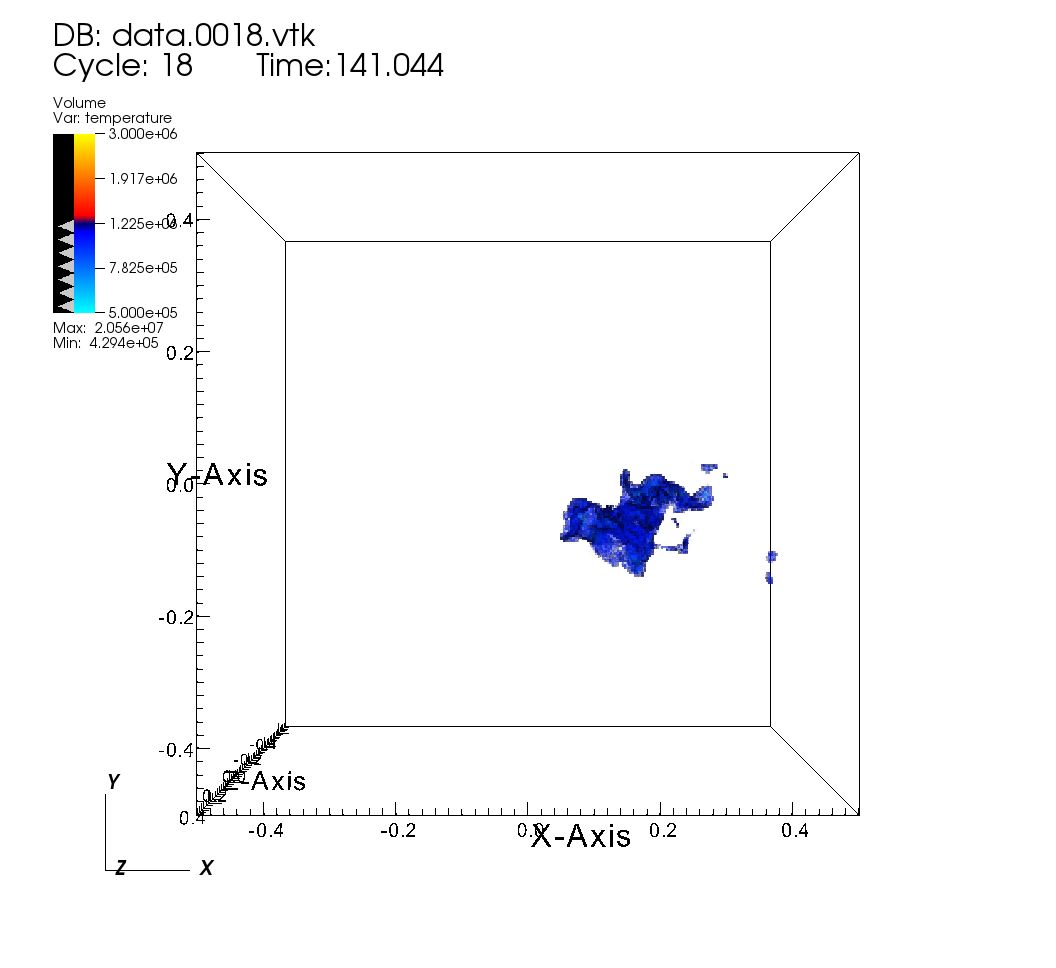} 
	} 
	\hfill
	\subfloat[]{ 
		\includegraphics[width=0.45\columnwidth]{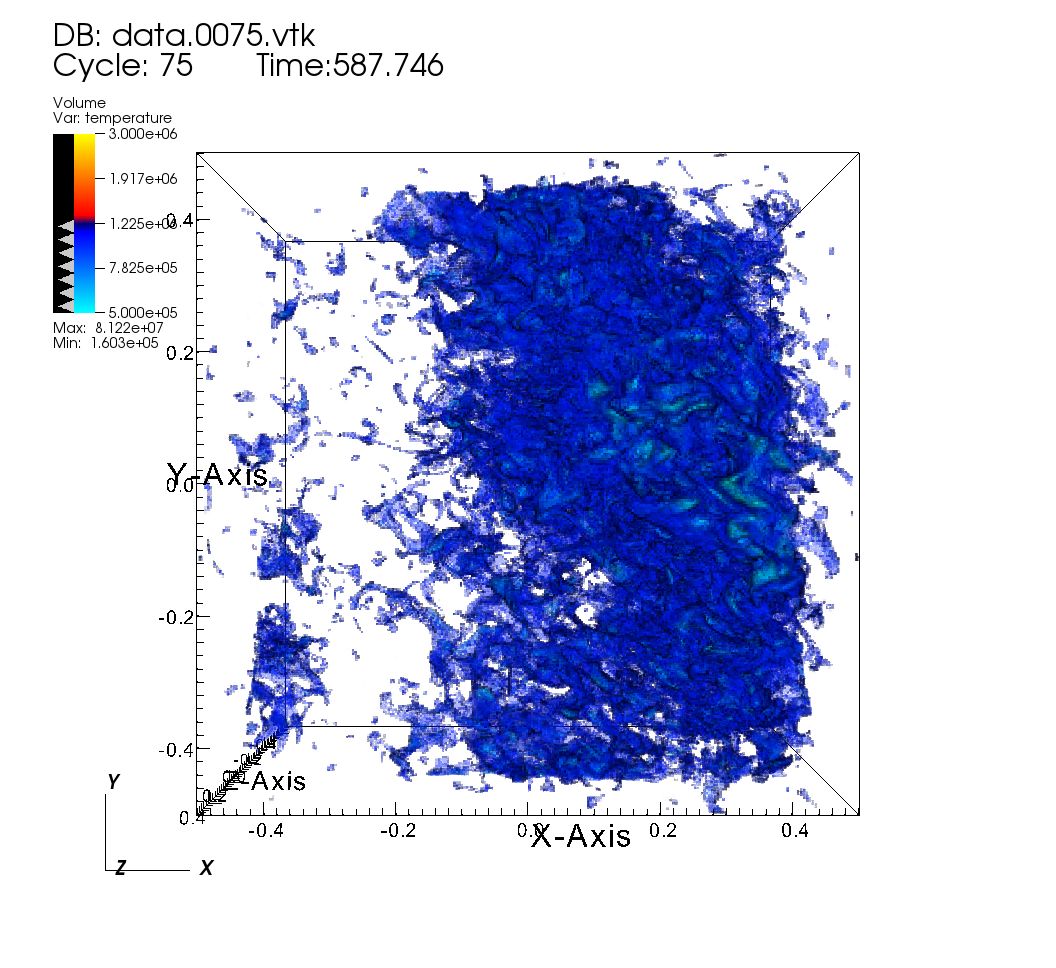} 
	} 
	\hfill 
	\caption[Cold gas volume rendering for initialized density perturbations]{Volume rendering of cold gas ($T<1.22\times10^6K$) for pure thermal heating run QD (upper panels) 
	and the run with equal thermal and turbulent heating BDh (lower panels). The left panels show the gas just after cold gas starts condensing, and the right panels represent 
	cold gas some time later. Note that the cloud shaped structures that form at later times for the QD run are almost at the same location as the initial clouds, denoting little gas 
	motion. There are cold filaments initially that eventually collapse on to the central core. 
	For BDh, the plots are similar to the Bh run without initial density perturbations (see lower panels of Fig. \ref{fig:B-l-h}).}\label{fig:D}
\end{figure}

\autoref{fig:del-rho-time} shows that multiphase gas in the high$-k$ driving run with initial perturbations (TDh) condenses out only slightly earlier ($\approx$1500 Myr) than the run 
without initial density perturbations (Th; $\approx 1700$ Myr). This is much longer than multiphase condensation without turbulence (QD), which happens on a cooling time. 
In fact, \autoref{fig:del-rho-time} shows that $\langle \delta \rho_{\rm rms} \rangle/\langle \rho \rangle$ is larger initially but attains the same amplitude as Th 
after $\approx 100$ Myr, suggesting that turbulence wipes out initially imposed isobaric density fluctuations on an eddy-turnover time. The run with high$-k$ driving
shows much gas below $T_{\text{cutoff}}$ and at intermediate temperatures (\autoref{fig:mach-temp-dist-D}). The temperature of the hottest gas is not as high as QD. The 
multiphase PDFs are similar to those of the Th run (\autoref{fig:mach-temp-dist-T}).

\subsubsection{Turbulent \& thermal heating (BDh)}

Initial density perturbations have a much bigger impact on the run with both turbulent and thermal heating (BDh; $f_{\text{turb}}=0.5$, $K_{\text{driving}}=12$) 
than with just turbulent heating (Th). \autoref{fig:del-rho-time} shows that cold gas condenses out for BDh at $t_{\text{mp}} \approx 400$ Myr, almost three times shorter than the 
run without initial density perturbations (Bh).  The shorter time scale of multiphase gas condensation is because of the decreased efficiency of turbulent mixing (since 
$f_{\text{turb}}=0.5$). Hence, the denser regions can cool to the stable temperature on a much shorter time scale. This time is still a factor of a few longer than 
$t_{\text{mp}}$ for pure thermal heating (QD).

\autoref{fig:mach-temp-dist-D} shows that the Mach number and temperature PDFs are qualitatively similar to the runs without density perturbations, but with a Mach 
number peak at $\approx 0.6$, somewhat lower than the pure turbulence run (Th; see  \autoref{fig:mach-temp-dist-T}). Volume rendering plots of density in \autoref{fig:D} (lower panels) 
are also qualitatively similar to the run Th (see \autoref{fig:T-l-h}) but with less mixing.

Most of the hot gas with partial thermal heating and initial density perturbations (BDh) is subsonic, and the timescale for multiphase condensation is not unrealistically long.
These properties match the observations qualitatively. In \autoref{subsec:disc-turb-heating} we further quantify the fraction of turbulent heating ($f_{\rm turb}$) by comparing 
with Hitomi observations.

\section{Discussion}
\label{sec:discussion}

This work has two key aims: (i) quantify the efficacy of unstratified turbulence in generating
density, pressure and surface brightness fluctuations; and 
(ii) quantify the extent to which turbulent heating can heat cool cores of clusters within the context of our idealized thermal balance simulations. 

In the first set of runs, we drive turbulence (mostly on large scales) with different forcing amplitudes and check the scaling of pressure, density and surface brightness fluctuations 
of the gas with the turbulent Mach number of the flow. We also calculate the power spectra of the same quantities and their variation with the wavenumber and Mach number.
In the second set of runs we impose thermal balance -- the sum of turbulent and thermal heating balances net cooling -- to mimic cluster cool cores. In some of these simulations 
we also drive turbulence at an order of magnitude 
smaller scale so that we get a smaller turbulent velocity ($\rho v^3 /l = \epsilon$, the energy input rate from turbulence; $v\propto l^{1/3}$ for the same $\epsilon$), close  to observations.

\subsection{Comparison with previous works}
\label{subsec:comparing_results}

We differ in two fundamental ways compared to the previous analyses (e.g., \citealt{gaspari2013constraining,zhuravleva2014relation,zhuravleva2014turbulent}) of this topic. First, we
do not include the background gravitational stratification and second, in our thermal balance setup cold gas can only form by condensation from the hot ICM. Both these assumptions have profound 
effects on our results and can essentially explain the seemingly different outcomes of our work compared to the previous studies. In the following paragraphs we motivate our choices
and highlight their impact on the outcomes of our study.

\citet{gaspari2013constraining} simulated hydro turbulence in the ICM of the Coma cluster and reported that $ \delta \rho/\rho  \propto \mathcal{M}_{\rm rms}$ even for subsonic driving.
This appears contradictory to the results from our fiducial simulations, but note that unlike us they use a stably stratified ICM. In a stably stratified atmosphere, turbulent driving 
can excite internal gravity waves for which the density perturbations are large relative to the pressure fluctuations ($\delta \rho/\rho \propto {\cal M}_{\rm rms} 
\gg \delta p/p$), and the power spectra are different from isotropic homogeneous turbulence (e.g., \citealt{lindborg2006}; see the recent book \citealt{verma2018}). 
Even for stably-stratified turbulence there seems to be a disagreement in the scaling of density and velocity power spectra. The high resolution simulations of \citet{kumar2014}
show the velocity and density power to be different ($\propto k^{-11/5}$ and $\propto k^{-7/5}$ respectively, in agreement with \citealt{bolgiano1959} but different from K41 scaling $\propto k^{-5/3}$ for both found by \citealt{gaspari2014}).
Thus, more work is needed to understand the relation between density and velocity fluctuations at different scales for parameters appropriate for galaxy clusters.

The ratio of the restoring buoyancy force and the nonlinear turbulent force can be defined as the scale-dependent turbulent Richardson number, 
\begin{equation}\label{eq:Richardson}
{\rm Ri}(l) = \frac{ \frac{g}{\gamma} \frac{d }{d\ln r} \ln \left( p/\rho^\gamma \right)  }{v^2(l)/l},
\end{equation}
 which is smaller at small scales ($l$) for K41 turbulence; i.e., turbulent force dominates over the buoyancy force at small scales 
 (\citealt{ruszkowski2010}). Here we assume the average vertical displacement to be the same as the size of the (isotropic) turbulent eddy ($l$). 
 If magnetized (anisotropic) conduction is of order the Spitzer value, the effective Richardson number is $\propto d\ln T/d\ln r$ and even smaller (\citealt{sharma2009b}). 
 
 Thus, for turbulent 
 velocities expected in both cool-core and non-cool-core clusters ($\gtrsim 100$ km s$^{-1}$) the effects of stratification may be small, especially at smaller scales. 
 Cosmological simulations of relaxed clusters (without cooling) agree with $\delta \rho/\rho \sim \mathcal{M}/\sqrt{3}$ scaling (e.g., see Figs. 2, 3 in \citealt{zhuravleva2014relation}), but this may break down at the smaller (10s of kpc) scales of cool cores where observations are probing below the Ozmidov scale (scale at which ${\rm Ri} \sim 1$; e.g., see the Extended Data Figure 4 in \citealt{zhuravleva2014turbulent}).  \citet{zhuravleva2014relation} argue that the scaling between the density and velocity fluctuations at small scales is inherited from the 
 buoyancy-dominated larger
 scales. This must be checked with high resolution simulations since kinetic energy flux crossing different $k$s is not expected to be a constant (unlike in K41) as it is converted into potential energy
 in a scale-dependent way. Moreover, turbulence is expected to be K41-like at small scales, irrespective of the behavior at large scales. 

Coming to our thermal balance simulations, note that the only way cold gas can be produced in these is via condensation from the hot phase through thermal instability in a medium
with global thermal balance. For this to happen, the turbulent mixing time of gas must be longer than the cooling time. This requirement puts an upper-limit on the turbulent velocity in our setup 
(see section \ref{subsec:disc-turb-heating}). However, if most cold gas in the ICM is due to other mechanisms, such as the uplifting of cold gas from the central galaxy, then $t_{\text{cool}}$ can 
be much longer than any other time scale, since cold gas does not condense out of the hot ICM. In \citet{zhuravleva2014turbulent} the cooling timescale is longer than the other relevant timescales 
because they do not assume the cold gas to condense out of the ICM. 
 
 \subsection{Adjusting $f_{\rm turb}$ to match Hitomi observations}\label{subsec:disc-turb-heating}

A necessary condition for the condensation of cold gas in a turbulent medium is that the turbulent mixing time be longer than the cooling time. However, in presence of gravity,
cold gas may not condense out even in absence of external turbulence if the ratio $t_{\rm cool}/t_{\rm ff} \gtrsim 20$ (\citealt{mccourt2012,choudhury2016}). In this regime, the 
amplitude of density perturbations is smaller for larger $t_{\rm cool}/t_{\rm ff}~(\gtrsim 20;$ e.g., see the right panel of Fig. 3 in \citealt{mccourt2012}). However, our
idealized set up without stratification is applicable for cool cluster cores with $t_{\rm cool}/t_{\rm ff} \lesssim 10$ in which multiphase gas is able to condense due to 
local thermal instability.

The ratio of the cooling time ($t_{\text{cool}}\equiv 1.5nk_BT/n_en_i\Lambda$, which is independent of length scale) and the turbulent mixing time 
($t_{\text{mix}}\equiv l/v_l$) is longer for smaller length scales $(v_l\propto l^{1/3}, \text{ }t_{\text{mix},l}\propto l^{2/3}$ for K41 turbulence). With thermal balance,
\begin{equation}
\label{eq:turb_cool_balance}
\dot{E}_{\text{turb}} \sim \rho v_l^2/t_{\text{mix}, l} \approx \rho v_l^3/l \approx f_{\text{turb}}\dot{E}_{\text{cool}} = f_{\text{turb}}U/t_{\text{cool}},
\end{equation}
where turbulent energy dissipation rate is scale independent, $f_{\text{turb}}$ is the turbulent heating fraction, and $U=P/(\gamma-1)$ is the thermal energy density. 
Thus, at the driving scale
\begin{equation}
\label{eq:tcool_by_tmix}
 t_{\text{cool}}/t_{\text{mix}, L} \approx f_{\text{turb}}U/2K \sim f_{\rm turb} {\cal M}^{-2}_{\rm rms},
\end{equation}
where $K=\rho v_L^2/2$ is the kinetic energy density at the driving scale ($L$). For smaller scales the ratio is longer (${\cal M}^{-2}_{\rm rms} [l/L]^{-2/3}$) and condensation is more difficult.

It is worth noting that the turbulent heating rate $\dot{E}_{\rm turb} \sim \rho v_L^3/L$ is very sensitive to $v_L$, and  can be matched with the average core cooling rate by only 
changing $v_L$ slightly (and $L$ to some extent). However, if the cold gas is to condense out of the hot phase due to thermal instability, the cooling time must be shorter than the turbulent mixing 
time. This constraints the Mach number in the hot phase to be $\gtrsim 1$ for turbulent heating to fully balance radiative losses ($f_{\rm turb}=1$; see Eq. \ref{eq:tcool_by_tmix}). For subsonic motions consistent with observations, 
turbulent heating fraction ($f_{\rm turb}$) needs to be small and/or turbulent driving must occur at small scales. This is what we argue next.

On scales ($l$) larger than the driving scale, turbulent diffusion happens due to eddies of size $L$ because energy only flows to smaller scale in K41 turbulence. 
The turbulent diffusion coefficient for $l \gtrsim L$ is given by 
$D_{\text{turb}}=Lu_L$ and the mixing time scale is
\begin{align}
t_{\text{mix},l>L} \sim l^2/D_{\text{turb}} \approx (l/L)^2t_{\text{mix},L}.
\end{align}
Thus, the condition for multiphase condensation due to thermal instability becomes
\begin{equation}
\label{eq:cool_mix}
t_{\text{cool}}/t_{\text{mix}} \approx f_{\text{turb}} (L/l)^2(U/2K) \sim  f_{\text{turb}} (L/l)^2 {\cal M}_{\rm rms}^{-2} < 1,
\end{equation}
which can be satisfied with $U \gtrsim 2K$ (or equivalently ${\cal M}_{\rm rms} \lesssim 1$) only for scales much larger than the driving scale ($l \gg L$) and/or for $f_{\rm turb} \ll 1$. Our simulation results
are consistent with this criterion. Large scale driving with $f_{\rm turb}=1$ (run Tl) indeed shows the Mach number peak in the hot phase at ${\cal M}_{\rm rms} > 1$ (see Fig. \ref{fig:mach-temp-dist-T}).
With driving at small scales (run Th), but still with $f_{\rm turb}=1$, the peak Mach number is smaller (${\cal M}_{\rm rms} \approx 1$), and the run with small scale driving and $f_{\rm turb}=0.5$ 
shows an even smaller Mach number peak (run Bh; see Fig. \ref{fig:mach-temp-dist-B}).

\begin{figure}
	\includegraphics[width=\columnwidth]{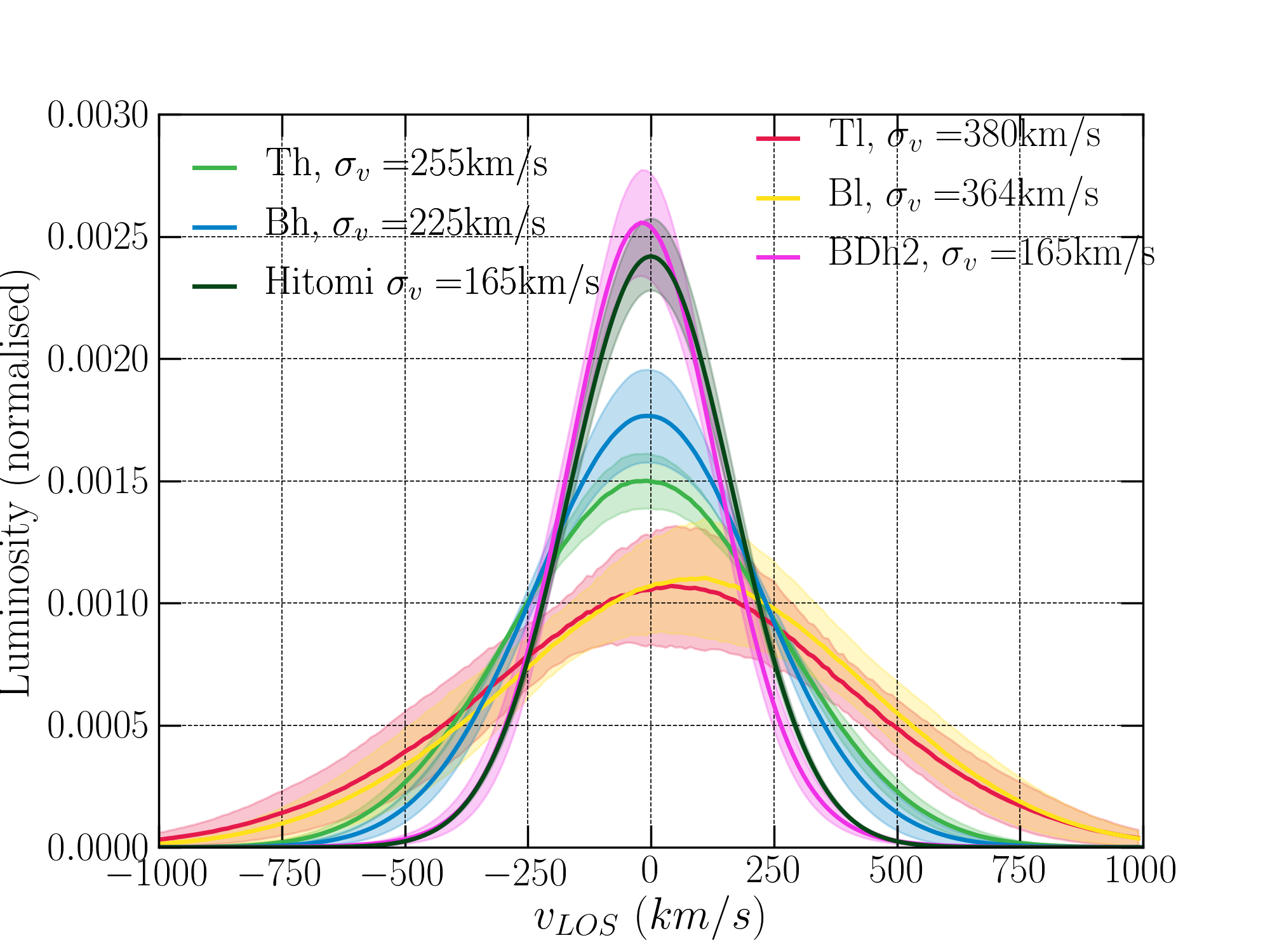}
	\caption[SB vs vlos PDF-heating+cooling runs]{Normalised PDF of X-ray luminosity versus the line of sight velocity ($v_{\text{LOS}}$) of the hot X-ray emitting gas ($T>5\times 10^6$ K) 
	for different thermal 
	balance runs. We calculate the luminosity and the LOS velocity of each grid cell in our simulation domain. Then we calculate the X-ray luminosity contributed within different $v_{\rm LOS}$ 
	bins. This PDF is a crude proxy for the X-ray lines which are broadened by turbulence in the hot ICM.
	The PDFs are well modelled by Gaussians, $\sigma_v$ being the standard deviation of the Gaussian. The solid line is the mean and the shaded region indicates $1-\sigma$ 
	variation in time after the condensation of multiphase gas. Even the run with small scale driving and $f_{\rm turb}=0.5$ (Bh) shows a much larger velocity dispersion as compared to the Hitomi 
	observations of Perseus core. The run with $f_{\text{turb}}=0.1$ and small scale driving (BDh2) produces close to the observed LOS velocity dispersion, with $\sigma_v=165$ km s$^{-1}$.}
	\label{fig:sb-vlos-pdf}
\end{figure}

The turbulent velocities for runs with $f_{\rm turb}=0.5$ (Bh, BDh; see Table \ref{tab:cool_runs}) are larger than what is measured by Hitomi observations of Perseus core. We therefore 
reduce turbulent forcing fraction ($f_{\rm turb}$) further to produce a line of sight velocity dispersion that is consistent with the observed value ($\approx 164$ km s$^{-1}$; see the last few rows 
in \autoref{tab:cool_runs}). Figure \ref{fig:sb-vlos-pdf}
shows the PDF of X-ray luminosity contributed at different line of sight (LOS) velocities for some of our thermal balance runs. We can produce the small LOS velocity dispersion measured by Hitomi
only with small turbulent heating ($f_{\rm turb} \approx 0.1$; run BDh2 in \autoref{tab:cool_runs}). 
ICM simulations with feedback AGN jets are also able to produce a velocity dispersion of similar magnitude for a substantial time, but it is more time variable than our 
idealized runs (\citealt{li2017,lau2017,prasad2018}). Another desirable feature of the run BDh2 is that cold gas
condenses out in a few cooling times (and not tens of cooling times as is the case for larger $f_{\rm turb}$ and small scale driving; e.g., runs Th, TDh, Bh in Table \ref{tab:cool_runs}).

\begin{figure}
	\subfloat[]{ 
		\includegraphics[width=0.45\columnwidth]{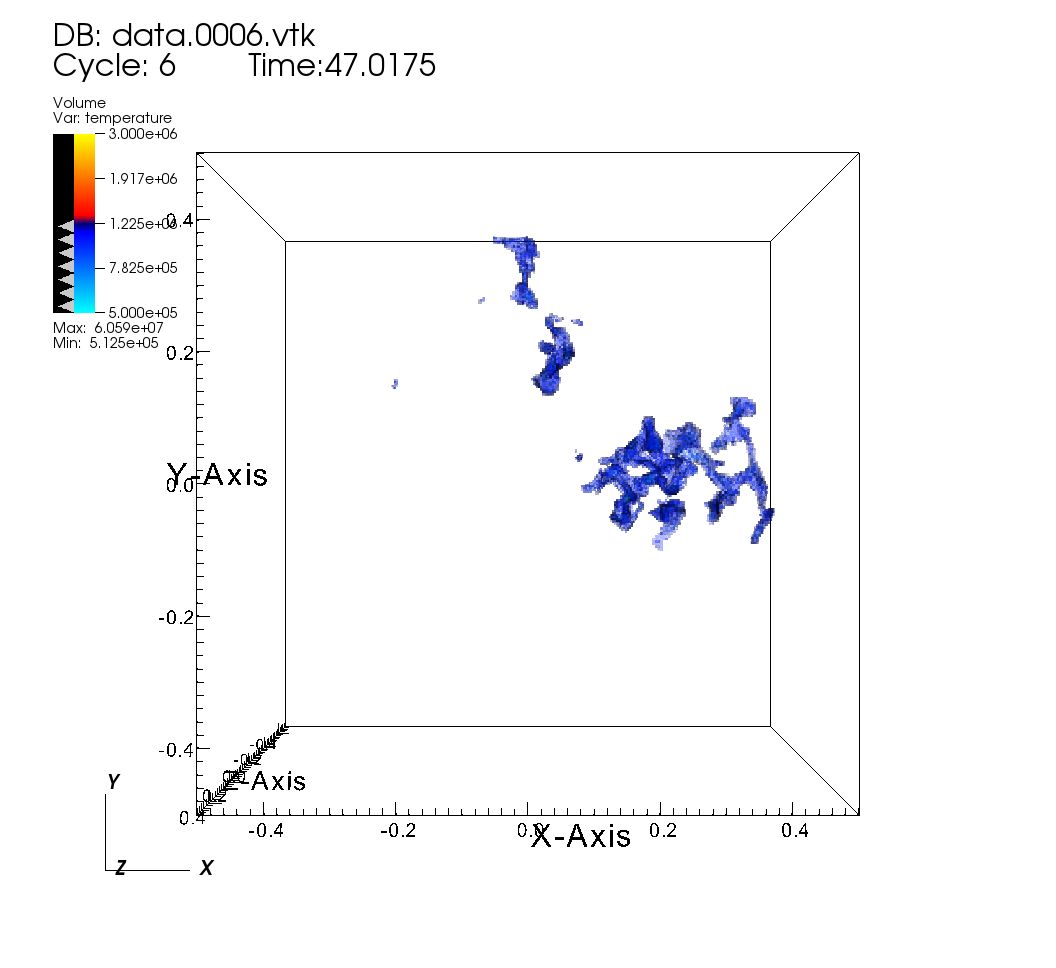} 
	} 
	\hfill 
	\subfloat[]{ 
		\includegraphics[width=0.45\columnwidth]{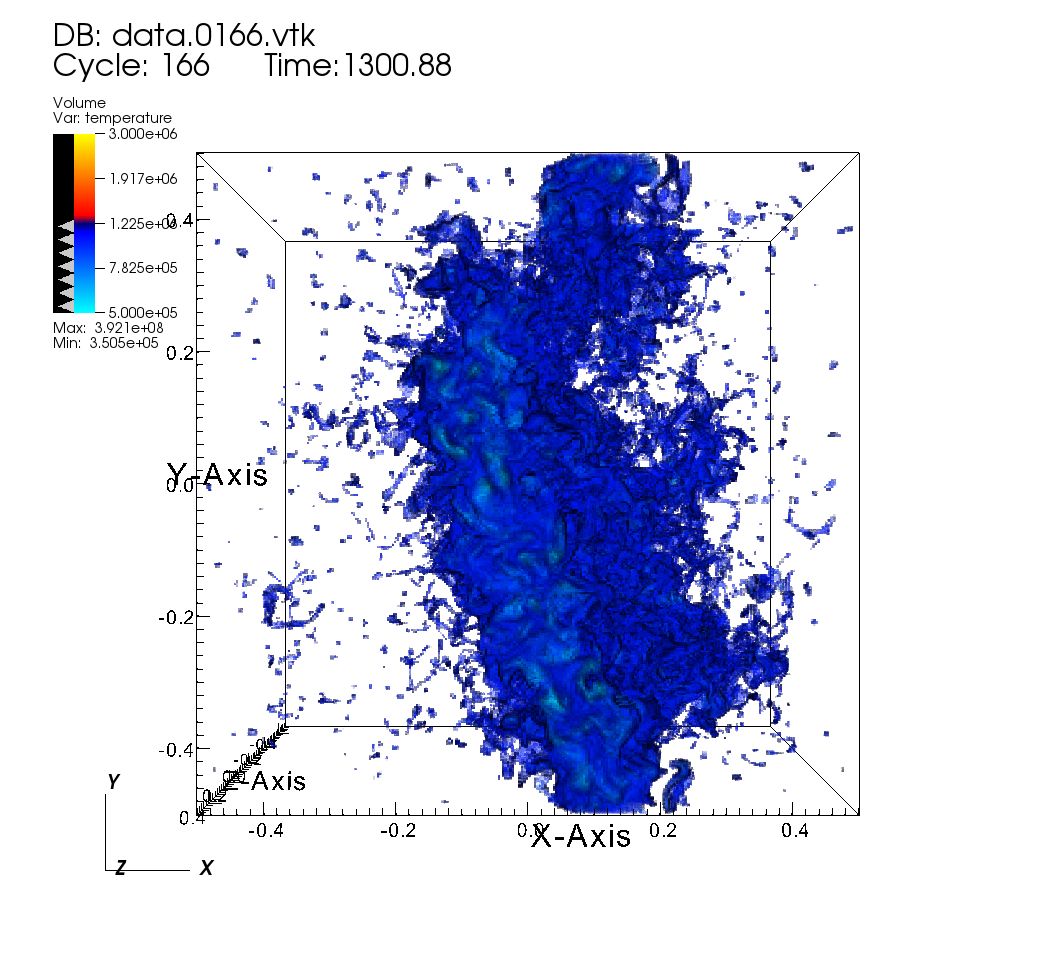} 
	} 
	\hfill 
	\subfloat[]{ 
		\includegraphics[width=\columnwidth]{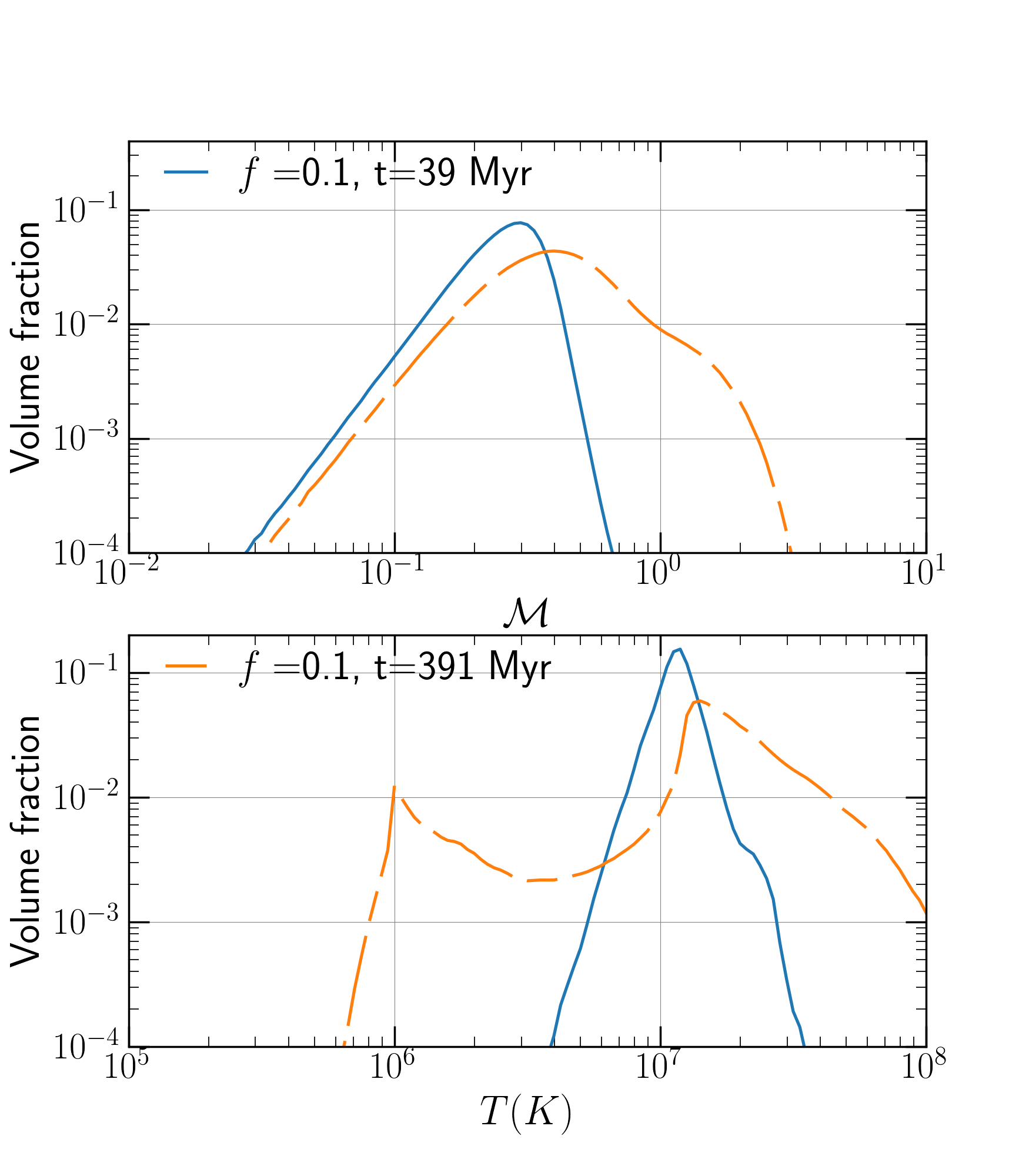} 
	} 
	\hfill
	\caption[Cold gas volume rendering for initialized density perturbations]{Upper panels: Volume rendering of cold gas ($T<1.22\times10^6K$) for the run BDh2 with $f_{\text{turb}}=0.1$ and initial 
	density perturbations. The left panel is a snapshot just after cold gas starts condensing, and the right panel is much later. The outer layers of the cloud
	look similar to BDh and TDh runs (runs with small scale turbulent driving), and not like the QD run with no turbulent forcing. 
	Middle and lower panels: Mach number and temperature PDFs before and after multiphase condensation for the same run. 
	The distribution is intermediate between that of QD and BDh runs. BDh2 has a Mach number peak at  $\mathcal{M} \approx 0.4$.}\label{fig:BDh2}
\end{figure}

The top panels of \autoref{fig:BDh2} show the cold gas volume rendering plot of our weak turbulence run ($f_{\rm turb}=0.1$; run BDh2) that matches Hitomi LOS velocity dispersion, 
just after condensation starts and later. The 
distribution of cold gas appears intermediate between pure turbulent heating runs (Fig. \ref{fig:T-l-h}) and pure thermal heating run (top panels of Fig. \ref{fig:D}). In particular, the cold gas cloud as a whole
appears stationary but its surface is turbulent. Of course, the addition of thermal conduction will wipe out small scale features in temperature and density of the hot phase 
(e.g., see Fig. 4 in \citealt{gaspari2014} and Fig. 1 in \citealt{wagh2014}), and anisotropic conduction makes the cold gas more filamentary (\citealt{sharma2010thermal}). 
The bottom two panels of \autoref{fig:BDh2} show the Mach number and temperature PDF for the same run at early and late times. The late time Mach number peak occurs at a
reasonable value of ${\cal M} \sim 0.4$ and the temperature of the hot phase peaks between 1 and 2 keV (a factor of two smaller than Perseus core so the comparison with 
Hitomi observations is not quantitative). The temperature distribution after condensation is bimodal with a lack of gas at intermediate temperatures.
 
\begin{figure}
	\includegraphics[width=\columnwidth]{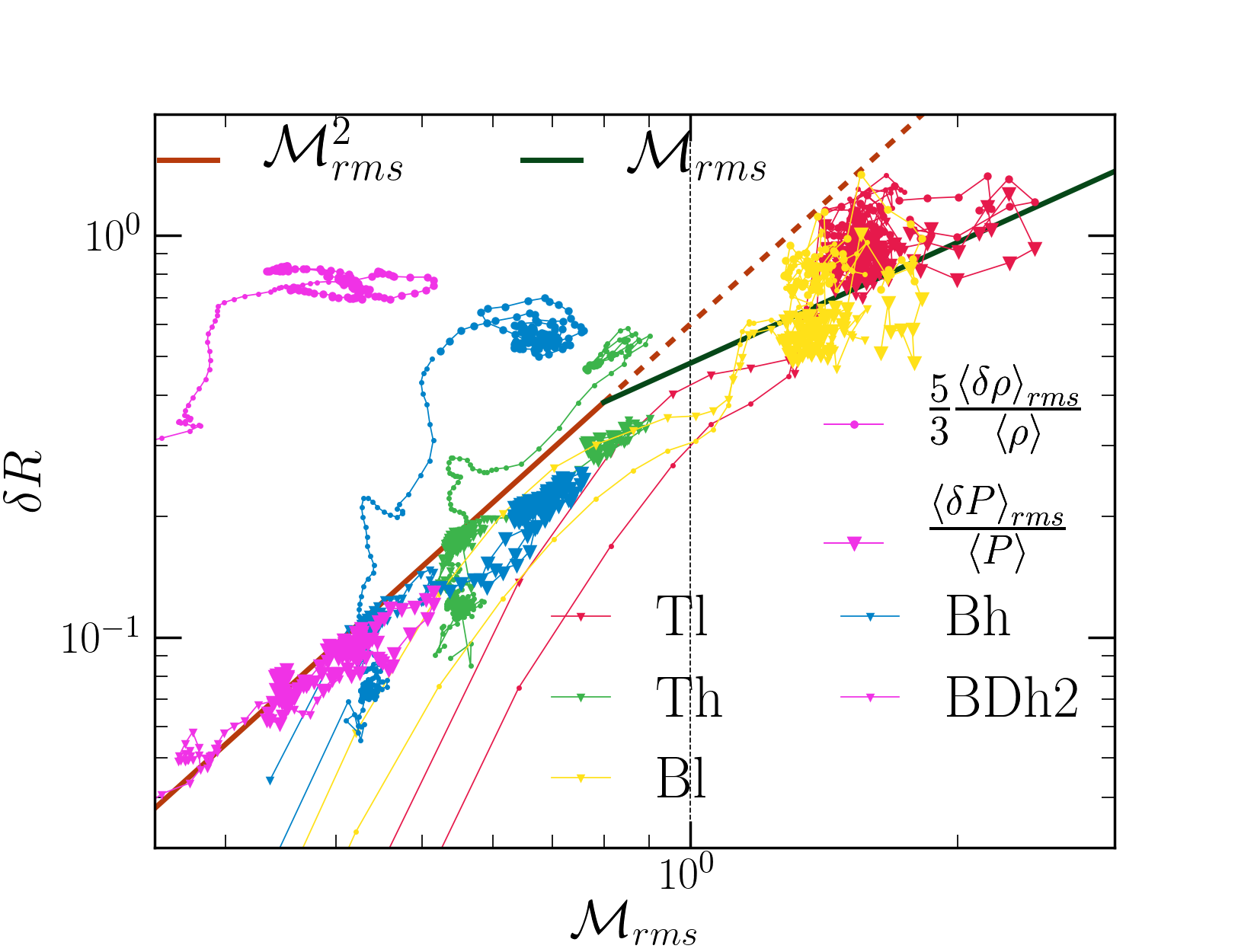}
	\caption[Density and pressure fluctuations vs Mach number-heating+cooling runs]{The root mean square (rms) density and pressure fluctuations of the hot X-ray emitting gas ($T>5\times 10^6$ K) 
	 as a function of the rms Mach number $\mathcal{M}_{\text{rms}}$ for some thermal balance runs. Data are plotted after 8 Myr in all cases.
	 The dark ${\cal M}_{\rm rms}^2$ and ${\cal M}_{\rm rms}$ lines are the turbulence-only scalings 
	 (Fig. \ref{fig:rho-mach}). 
	 These graphs show both the turbulent steady state before condensation (with smaller markers) and the state after condensation (with larger markers). The evolution
	 is qualitatively different from pure turbulence runs in which the rms Mach number and density/pressure fluctuations decrease with time because of heating. Here the density fluctuations are much
	 higher than the turbulent scaling, especially at small $\mathcal{M}_{\text{rms}}$. In fact, the rms density perturbations after multiphase condensation are similar for different runs (see also 
	 Fig. \ref{fig:del-rho-time}). Note, however, that the pressure fluctuations follow the scalings from the turbulent runs even after condensation.}	
	 \label{fig:rho-prs-mach-cool}
\end{figure}
\begin{figure}
	\includegraphics[width=\columnwidth]{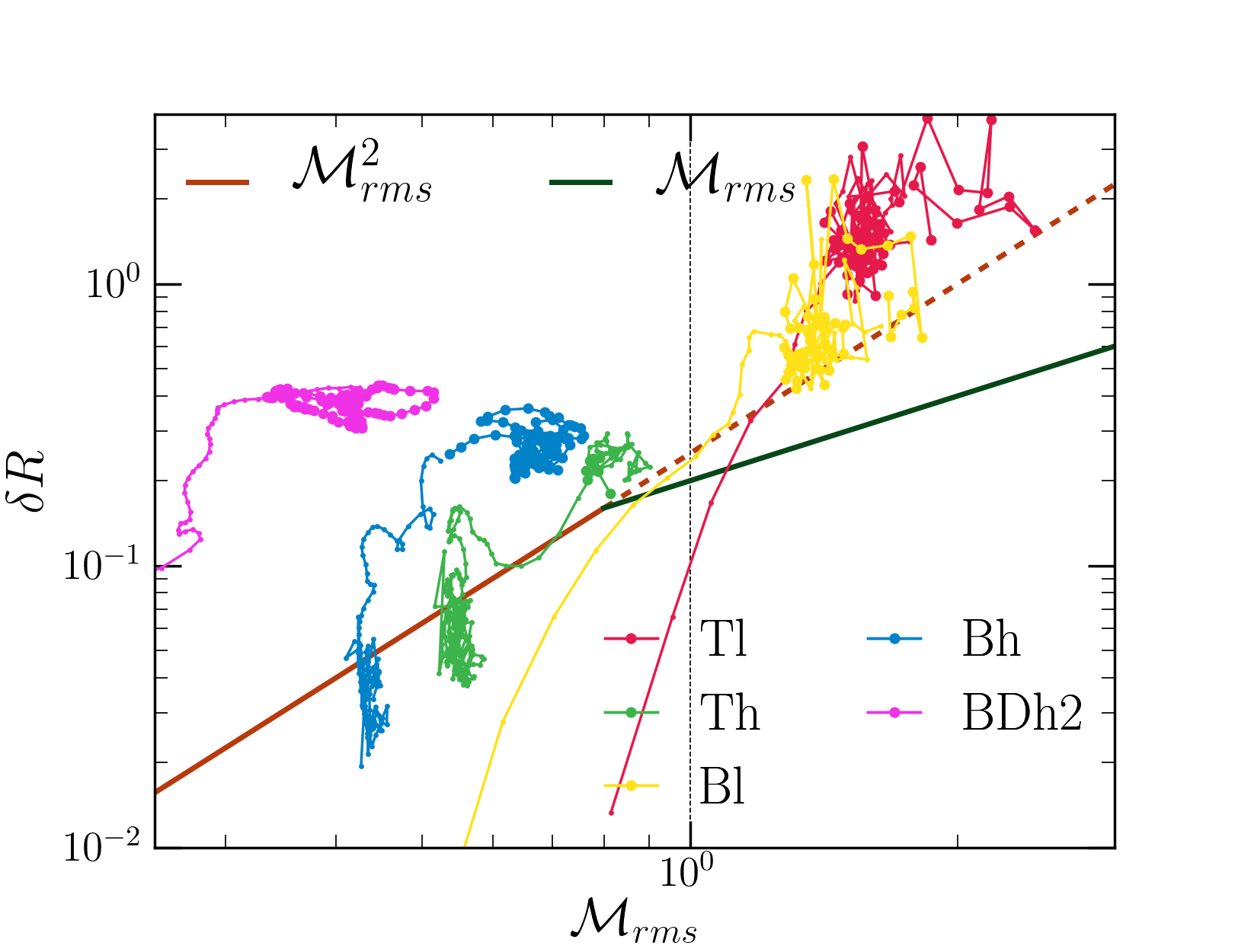}
	\caption[Surface brightness fluctuations vs Mach number-heating+cooling runs]{The rms surface brightness (SB) fluctuations of hot gas ($T>5\times 10^6$ K) as a 
	function of the rms Mach number $\mathcal{M}_{\text{rms}}$ for some thermal balance runs. Data are plotted after 8 Myr in all cases. 
	The  ${\cal M}_{\rm rms}^2$ and ${\cal M}_{\rm rms}$ fits are from the
	turbulence-only runs (Fig. \ref{fig:sb-mach}). Again, SB fluctuations are much larger than the scaling with only turbulence. From Fig. \ref{fig:rho-prs-mach-cool}, the projected 
	pressure fluctuations (not shown) are expected to follow the pure turbulence scaling. The supersonic 
	runs clearly show very large rms SB fluctuations than the  turbulence-only runs. The thicker (thinner) markers are for times after (before) multiphase condensation.}
	\label{fig:sb-mach-cool}
\end{figure}

\subsection{Scaling of density, pressure \& surface brightness perturbations}
\label{subsec:disc-scaling-sb-rho-vel}

Most of the work relating density and surface brightness fluctuations (measured from X-ray observations) in the ICM to the level of turbulence has not included the effects of cooling and heating.
While this is justifiable for non-cool core clusters and for cluster outskirts that have long cooling times, the cool cores are fundamentally affected by cooling and heating. Our idealized thermal balance 
runs (see \autoref{sec:ResultsHC}) are a step towards making the cool-core turbulence models more realistic. In fact, the density perturbations because of local thermal instability, which can lead to
multiphase gas, are much larger than what is expected from K41 turbulence. The caveat, however, is that we do not include the cluster gravity which can suppress condensation and density 
fluctuations to some extent.

\autoref{fig:rho-prs-mach-cool} shows the rms density and pressure fluctuations as a function of the rms Mach number for some of our thermal balance runs. The thick solid lines show the
scaling from pure turbulence runs (Fig. \ref{fig:rho-mach}). The density fluctuations are much larger than pure turbulence because isobaric (because sound crossing time is shorter than 
cooling time) thermal instability leads to large density fluctuations but not large turbulent velocities. Note that density perturbations are large {\em even before} condensation. 
Not only are the density fluctuations much larger than the scaling for 
isotropic/homogeneous turbulence, they are also larger than the linear extrapolation of supersonic scaling or scaling of density perturbations with internal gravity waves in a stratified atmosphere
($\langle \delta \rho/\rho \rangle_{\rm rms} \sim {\cal M}_{\rm rms}/\sqrt{3}$). 

Similarly, in \autoref{fig:sb-mach-cool} the surface brightness fluctuations are also much larger.
The implication is that the turbulent velocities inferred from density fluctuations can be much higher if thermal instability is ignored (as in \citealt{zhuravleva2014turbulent}). 
The pressure fluctuations, in contrast, are smaller and consistent with isotropic/homogeneous
turbulence (Fig. \ref{fig:rho-prs-mach-cool}). Also note that the pressure fluctuations for subsonic internal gravity waves are much smaller than density fluctuations. \autoref{tab:fluctuations} lists
the nature of density, pressure and velocity perturbations for different regimes relevant to the ICM. Future comparison of X-ray surface brightness maps (from Chandra/XMM maps), Sunyaev-Zeldovich 
fluctuations (which probe the line of sight pressure fluctuations) and turbulent broadening in X-ray lines (e.g., by successors of Hitomi) can teach us much about the nature of dominant 
fluctuations in the ICM.

\begin{table}
\caption{Nature of fluctuations}
{%
\begin{tabular}{c c }
\hline
Fluctuations  &  for ${\cal M}_{\rm rms}<1$ \\
\hline
isotropic/homogeneous turbulence &  $\delta p/p \sim (5/3)(\delta \rho/\rho) \sim {\cal M}_{\rm rms}^2$  \\ 
internal gravity waves  & $\delta p/p \sim {\cal M}_{\rm rms}^2$, $\delta \rho/\rho \sim {\cal M}_{\rm rms}$ \\ 
thermal instability + turbulence  & $\delta p/p \sim {\cal M}_{\rm rms}^2$, $\delta \rho/\rho > {\cal M}_{\rm rms}$ \\ 
\hline
\end{tabular}}
\label{tab:fluctuations}
\end{table}

\begin{figure}
	\includegraphics[width=\columnwidth]{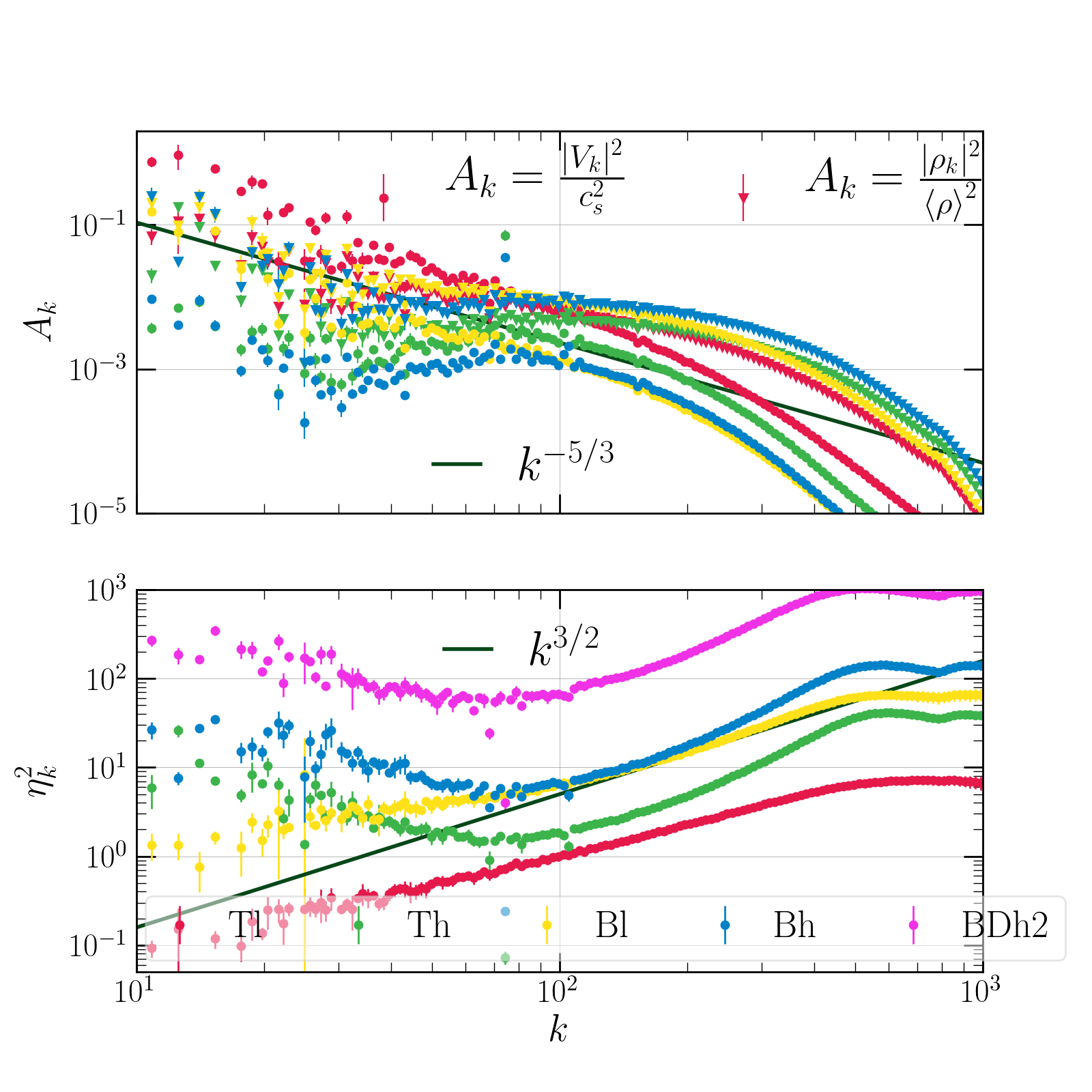}
	\caption{The normalized density and velocity power spectra (top panel) and their ratio (bottom panel) for some thermal balance runs. Compared to pure turbulence runs (see Fig. \ref{fig:rho-vel-spectra}),
	the density power spectrum is much larger and shallower than the velocity power spectrum, with their ratio ($\eta_k^2$) increasing with the wavenumber as 
	$k^{3/2}$ in the inertial range. In other words, the density power spectrum scaling with thermal balance is close to $k^{-1/6}$.
	}	
	 \label{fig:rho-vel-cool-spectra}
\end{figure}

\begin{figure}
	\includegraphics[width=\columnwidth]{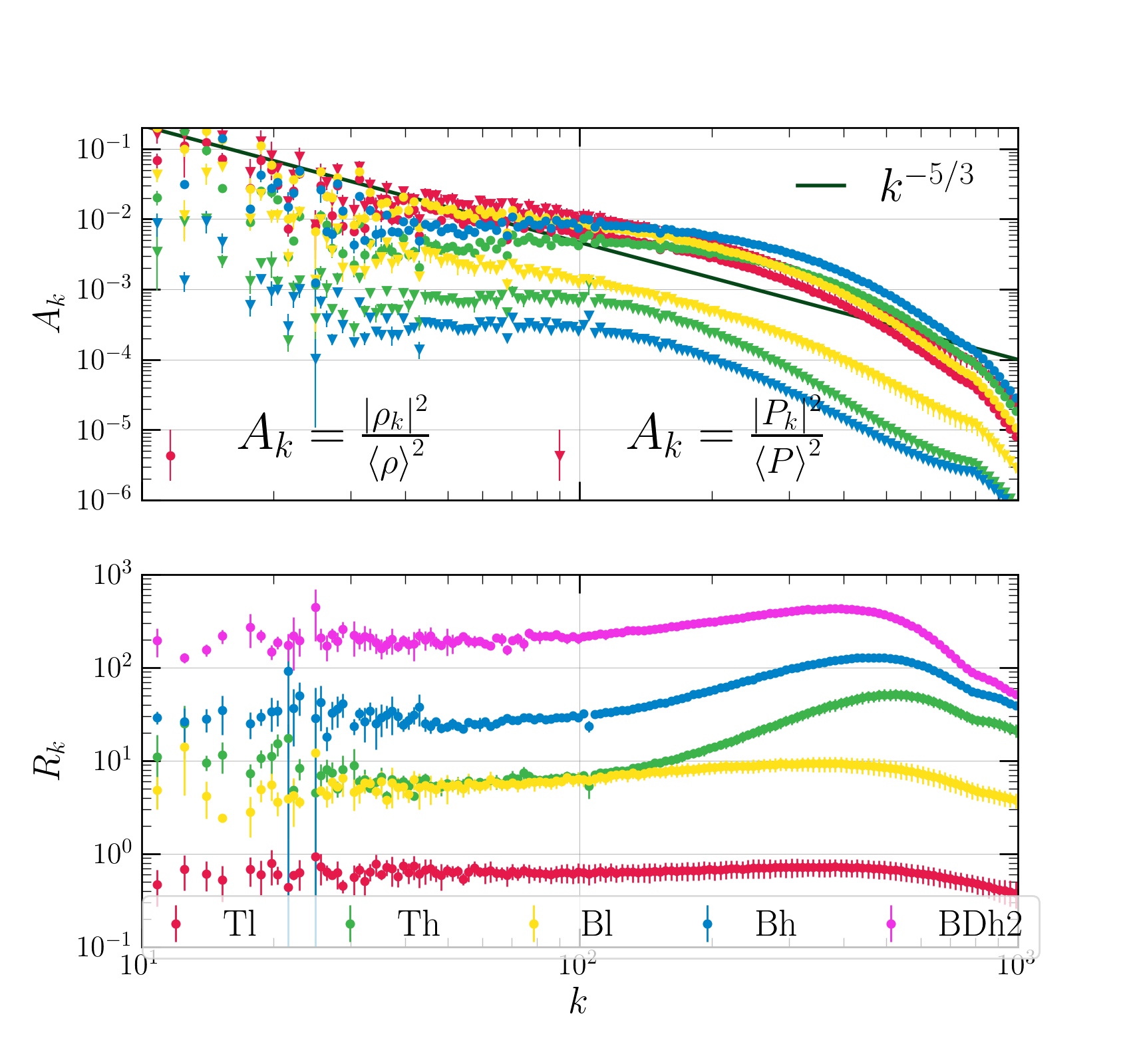}
	\caption{The normalized density and pressure power spectra (top panel) and their ratio (bottom panel) for some thermal balance runs. 
	Compared to pure turbulence runs (see the top panel of Fig. \ref{fig:rho-vel-spectra}),
	the density power spectrum is much larger and shallower than the pressure power spectrum. The ratio of density and pressure power spectra 
	($R_k$ in the bottom panel) is almost constant in the supersonic regime; i.e., $|P_k|/\langle P \rangle \approx (5/3) |\rho_k|/\langle \rho\rangle$ still holds in the supersonic 
	regime (run Tl) with thermal balance. In the subsonic regime, pressure power spectrum scaling even with cooling and heating is similar to the velocity power spectrum, 
	$\propto k^{-5/3}$. 
	}
	\label{fig:rho-prs-cool-spectra}
\end{figure}

Figures \ref{fig:rho-vel-cool-spectra} and \ref{fig:rho-prs-cool-spectra} show the density, pressure and velocity power spectra as a function of wavenumber ($k$). The amplitude of density
fluctuations is much higher (as expected from \autoref{fig:rho-prs-mach-cool}) and the density fluctuation spectrum is much shallower than K41 spectrum (expected in absence of 
cooling/heating). The velocity and pressure power spectra in the subsonic regime are consistent with K41 turbulence (see Fig. \ref{fig:rho-vel-spectra}). A similar nature for the power spectra
is also seen for the pure thermal heating run QD (not shown in these figures). 


\section{Conclusions}\label{sec:conclusion}
We have carried out high resolution simulations of turbulence relevant to the intracluster medium (ICM), and analysed scaling of various physical quantities and observables. 
Unlike most previous works, we explicitly consider the influence of cooling and heating in the cluster core on density, pressure and velocity fluctuations. Based on our simulations, 
following are our key conclusions.

\begin{itemize}
	\item The turbulent heating rate $\dot{E}_{\rm turb} \sim \rho v_L^3/L$ is very sensitive to $v_L$, and  can be matched with the average core cooling rate by changing $v_L$ 
	slightly (and to some extent by changing $L$; e.g., see section 6.2 of \citealt{zhuravleva2018}).However, if the cold gas is to condense out of the hot phase 
	due to thermal instability, the cooling time must be shorter than the turbulent mixing time.
	This constraints the Mach number in the hot phase to be $\gtrsim 1$ for driving on 10s of kpc, which is ruled out by observations. Driving at smaller scales somewhat reduces 
	the Mach number in the hot phase ($v_L \propto L^{1/3}$; see Eq. \ref{eq:turb_cool_balance}), but it is still much larger than observations. Moreover, small-scale driving
	delays cold gas condensation because of short mixing time on the driving scale. In the context of our thermal balance models with multiphase condensation, 
	the only satisfactory way of matching the
	turbulent velocity measured by Hitomi in the core of Perseus cluster is by reducing the fraction of turbulent heating to $\sim 0.1$ of the cooling rate (see \autoref{subsec:disc-turb-heating}). 
	Thus, turbulent heating
	does not seem to be the dominant heating source in cool cores. Other sources that do not contribute much fluid motion in the hot phase provide $\sim 90\%$ of the feedback
	heating. Turbulent heating fraction is even smaller for driving at larger scales.  
	Also with cooling present, density and surface brightness fluctuations due to local thermal instability can be much larger than what is anticipated from turbulence-driven internal gravity waves. 
		
	\item The ratio $\eta_k^2$ between the density and velocity power spectra is much higher and scale dependent for thermal balance runs (see \autoref{subsec:disc-scaling-sb-rho-vel}). 
	For pure isotropic/homogeneous turbulence in the subsonic regime, this ratio is independent of scale ($k$) but increases linearly with the Mach number (see Fig. \ref{fig:rho-vel-spectra}).
	For comparison, this ratio $\eta_k$ seems to be close to $1/\sqrt{3}$ and independent of $k$ when the background stable stratification is important, and cooling and heating 
	are ignored (\citet{gaspari2013constraining,zhuravleva2014relation}; see section \ref{subsec:comparing_results}).
	
	\item For thermal balance simulations, the density and surface brightness (SB) fluctuations are much larger than their scaling with the Mach number for turbulence simulations, and even
	compared to the density fluctuations seeded by internal gravity waves (see Table \ref{tab:fluctuations} \& \autoref{subsec:disc-scaling-sb-rho-vel}). Matching the X-ray surface brightness 
	fluctuations with turbulence or gravity wave scaling would lead to an overestimate of turbulent velocities. The power spectrum of density with heating/cooling is much larger and 
	shallower compared to K41 scaling, but the pressure power spectrum is similar to the velocity power spectrum, which follows K41 $k^{-5/3}$ scaling 
	(Figs. \ref{fig:rho-vel-cool-spectra}, \ref{fig:rho-prs-cool-spectra}). Thus, comparing X-ray surface brightness,
	high resolution spectra of X-ray lines, and the fluctuations of the Sunyaev-Zeldovich signal can tell us about the nature of perturbations in the ICM.

\end{itemize}

An important caveat 
of our simulations is that we do not include gravitational stratification, so internal gravity waves that can be excited by turbulence and lead to density fluctuations are absent. Although 
stratification is weak in galaxy clusters, it is necessary to include it in combination with cooling and heating to draw firm conclusions about the nature of fluctuations in cluster cores.
These fluctuations are a treasure-trove of information about physical processes in the ICM.

\section*{Acknowledgements}

We acknowledge the high performance computing facilities (in particular SahasraT) at the Supercomputer Education and Research Centre (SERC), Indian Institute of Science. This work was partially 
supported by an India-Israel joint research grant (6-10/2014[IC]) and a Swarnajayanti Fellowship from the Department of Science and Technology (DST/SJF/PSA-03/2016-17). 
PS thanks the Humboldt Foundation to enable his sabbatical at MPA. PS acknowledges very helpful discussions with Eugene Churazov. 
RM thanks Naveen Yadav for his valuable inputs during the start of this project. RM thanks Christoph Federrath for his valuable suggestions on the paper. We thank Xun Shi and Noam Soker for
helpful email exchanges and the referee Irina Zhuravleva for a constructive report.



\section*{Additional links}
The movies for the evolution of cold gas in different simulations  in the paper is available at: \url{https://www.mso.anu.edu.au/~rajsekha/BT_movies.html}.

\bibliographystyle{mnras}
\bibliography{refs}

\appendix

\section{Computing power spectra}
\label{sec:appdx_pow_spectra}
Since we use a discrete 3-D grid, the Fourier transform $A_k(\mathbf{k})$ is obtained by taking a discrete Fourier transform (DFT) of the real space data $A(\mathbf{r})$,
\begin{equation}
A_k(\mathbf{k})=\sum_{\mathbf{r}} A(\mathbf{r})e^{-\iota\mathbf{k}\cdot\mathbf{r}},
\end{equation}
where each component of $\mathbf{k}$ takes a values $[-\pi N/L, -\pi(N-1)/L, ..., \pi(N-1)/L, \pi N/L]$ ($L$ is the box size and $N$ the number of grid points in each direction) along the three directions.
We can create spherical shells in $k-$ space and define the power spectrum $E_k(k)$ as
\begin{align}
E_k(k)\Delta k&=\sum_{k\leq|\mathbf{k}|<k+\Delta k}|A_k(\mathbf{k})|^2,\\
\text{or }E_k(k)&=\sum_{k\leq|\mathbf{k}|<k+\Delta k}\frac{|A_k(\mathbf{k})|^2}{\Delta k},
\end{align}
where $\Delta k$ is the bin size. 

Since we have a large range of $k$s, we use a uniformly spaced grids in $\ln k$, with 
\begin{equation}
\Delta  \ln k =\frac{1}{n_{\text{bin}}}\ln \left(\frac{k_{\text{max}}}{k_{\text{min}}}\right),
\end{equation}
where $n_{\text{bin}}$ is the number of bins into which we divide the $k$-space, and $k_{\text{max}}$ and $k_{\text{min}}$ are the maximum and minimum wave numbers given by
\begin{align}
k_{\text{max}}&=\frac{2N\pi}{L},\\
k_{\text{min}}&=\frac{2\pi}{L}.
\end{align}
Note that $k_{\text{max}} > \sqrt{3} N \pi/L$, the maximum value of $|k|$. So the $i^{\text{th}}$ bin-boundary is given by
\begin{equation}
k_{\text{bin, i}}=k_{\text{min}} \left( \frac{k_{\text{max}}}{k_{\text{min}}} \right)^{i/n_\text{bin}},
\end{equation}
with $i=0,...,n_\text{bin}$, and
\begin{equation}
\Delta k_{\text{bin}, i} \equiv k_{\text{bin}, i} - k_{\text{bin}, i-1} = k_{{\rm bin}, i} \left( 1- \left[ \frac{k_\text{min}}{k_\text{max}} \right]^{1/n_{\rm bin}} \right).
\end{equation}
The power spectrum is then given by
\begin{equation}
E_k(k_i)=\sum_{k\leq|\mathbf{k}|<k+\Delta k_{\text{bin, i}}}\frac{|A_k(\mathbf{k})|^2}{\Delta k_{\text{bin, i}}}.
\end{equation}

\section{Calculating density and surface brightness spectra}
\label{sec:appdx_surf_brightness}
The central region of a cluster is its brightest part and is the major contributor to the surface brightness profile of the cluster. In our simulations, we model the central core of a cluster. This region is roughly spherical. 
However, we model it in a 3D Cartesian setup. So, while calculating the density and surface brightness power spectra, we use a roughly spherical density profile given by
\begin{subequations}
	\begin{align}
	&\delta\rho(\mathbf{r})=\rho(\mathbf{r})-\rho_0\\
	&\rho^\prime(\mathbf{r}) = \rho_0+\frac{\delta \rho(\mathbf{r})}{2}\left[1-\tanh\left(\frac{|\mathbf{r}|-|\mathbf{r_0}|}{\sigma}\right)\right], 
	\end{align}
\end{subequations}
where $\rho_0$ is the mean density, $\rho(\mathbf{r})$ is the density at a given location in our simulations, and $\rho^\prime(\mathbf{r})$ is the modified spherical density that we use for calculating the 
power spectrum of surface brightness. The transition scale of density perturbations, $\sigma$, has been set to $0.2L$, where $L$ is the size of our Cartesian box.

The weighting function decreases smoothly from one at $\mathbf{r}=0$ to zero at around $|\mathbf{r}|=|\mathbf{r_0}|$.  The 2-D surface brightness map is given by
\begin{equation}
SB(x,y)=\int_{-L/2}^{L/2}n^{\prime 2}(x,y,z)\Lambda(T)dz,
\end{equation}
where $n^\prime$ = $\rho^\prime/(\mu m_p)$.
This analysis method only affects the low-$k$ (large scale) modes of the surface brightness power spectrum. The inertial range remains unaffected by this method.

\bsp	
\label{lastpage}
\end{document}